\documentclass[pdflatex]{sn-jnl}

\usepackage{graphicx}%
\usepackage{amsmath,amssymb,amsfonts, mathtools}%
\usepackage{amsthm}%
\usepackage[title]{appendix}%
\usepackage{xcolor}%
\usepackage{booktabs}%
\usepackage{algorithm}%
\usepackage{algorithmicx}%
\usepackage{algpseudocode}%
\usepackage{float}
\usepackage{cleveref}
\usepackage{multirow}
\usepackage{subcaption}
\usepackage{xspace}
\usepackage{bm}
\usepackage{dsfont}
\usepackage{pifont}
\usepackage{enumitem}

\theoremstyle{thmstyleone}%
\theoremstyle{thmstyletwo}%

\theoremstyle{thmstylethree}%

\newcommand{\xmark}[1][red]{\textcolor{#1}{\ding{54}}}

\definecolor{TrollRed}{HTML}{E41A1C}
\definecolor{TrollTeal}{HTML}{1B9E77}
\definecolor{TrollCyan}{HTML}{17BECF}

\providecommand{\theHsection}{}
\providecommand{\theHsubsection}{}
\providecommand{\theHsubsubsection}{}
\providecommand{\theHfigure}{}
\providecommand{\theHtable}{}
\providecommand{\theHequation}{}
\providecommand{\theHalgorithm}{}

\newcommand{\startsupplement}{%
  \clearpage
  \setcounter{section}{0}%
  \setcounter{subsection}{0}%
  \setcounter{subsubsection}{0}%
  \setcounter{figure}{0}%
  \setcounter{table}{0}%
  \setcounter{equation}{0}%
  \setcounter{algorithm}{0}%
  \renewcommand{\thesection}{S\arabic{section}}%
  \renewcommand{\thesubsection}{\thesection.\arabic{subsection}}%
  \renewcommand{\thesubsubsection}{\thesubsection.\arabic{subsubsection}}%
  \renewcommand{\thefigure}{S\arabic{figure}}%
  \renewcommand{\thetable}{S\arabic{table}}%
  \renewcommand{\theequation}{S\arabic{equation}}%
  \renewcommand{\thealgorithm}{S\arabic{algorithm}}%
  \renewcommand{\theHsection}{supp.\arabic{section}}%
  \renewcommand{\theHsubsection}{supp.\arabic{section}.\arabic{subsection}}%
  \renewcommand{\theHsubsubsection}{supp.\arabic{section}.\arabic{subsection}.\arabic{subsubsection}}%
  \renewcommand{\theHfigure}{supp.\arabic{figure}}%
  \renewcommand{\theHtable}{supp.\arabic{table}}%
  \renewcommand{\theHequation}{supp.\arabic{equation}}%
  \renewcommand{\theHalgorithm}{supp.\arabic{algorithm}}%
  \crefname{section}{Section}{Sections}%
  \Crefname{section}{Section}{Sections}%
  \crefname{subsection}{Section}{Sections}%
  \Crefname{subsection}{Section}{Sections}%
  \crefname{figure}{Figure}{Figures}%
  \Crefname{figure}{Figure}{Figures}%
  \crefname{table}{Table}{Tables}%
  \Crefname{table}{Table}{Tables}%
  \crefname{equation}{Equation}{Equations}%
  \Crefname{equation}{Equation}{Equations}%
  \crefname{algorithm}{Algorithm}{Algorithms}%
  \Crefname{algorithm}{Algorithm}{Algorithms}%
}

\begin{document}

\title[Article Title]{Beyond Content: Behavioral Policies Reveal Actors in Information Operations}


\author*[1]{\fnm{Philipp J.} \sur{Schneider}}\email{philipp.schneider@epfl.ch}
\equalcont{These authors contributed equally to this work.}

\author[2]{\fnm{Lanqin} \sur{Yuan}}\email{lanqin.yuan@student.uts.edu.au}
\equalcont{These authors contributed equally to this work.}

\author[2]{\fnm{Marian-Andrei} \sur{Rizoiu}}\email{marian-andrei.rizoiu@uts.edu.au}

\affil*[1]{\orgdiv{College of Management of Technology}, \orgname{\'{E}cole Polytechnique F\'{e}d\'{e}rale de Lausanne}, \orgaddress{\postcode{CH-1015} \city{Lausanne}, \country{Switzerland}}}

\affil[2]{\orgdiv{Faculty of Engineering and Information Technology}, \orgname{University of Technology Sydney}, \orgaddress{\city{Ultimo} \state{NSW} \postcode{2007}, \country{Australia}}}

\abstract{
The detection of online influence operations---coordinated campaigns by malicious actors to spread narratives---has traditionally depended on content analysis or network features. These approaches are increasingly brittle as generative models produce convincing text, platforms restrict access to behavioral data, and actors migrate to less-regulated spaces. We introduce a platform-agnostic framework that identifies malicious actors from their behavioral policies by modeling user activity as sequential decision processes. We apply this approach to 12,064 Reddit users, including 99 accounts linked to the Russian Internet Research Agency in Reddit's 2017 transparency report, analyzing over 38 million activity steps from 2015-2018.
Activity-based representations, which model how users act rather than what they post, consistently outperform content models in detecting malicious accounts. When distinguishing trolls---users engaged in coordinated manipulation---from ordinary users, policy-based classifiers achieve a median macro-F1 of 94.9\%, compared to 91.2\% for text embeddings. Policy features also enable earlier detection from short traces and degrade more gracefully under evasion strategies or data corruption. These findings show that behavioral dynamics encode stable, discriminative signals of manipulation on Reddit's IRA-linked campaign, and point to resilient detection strategies in the era of synthetic content and limited data access.
}



\maketitle

\section*{Introduction}
\label{sec:introduction}

Coordinated efforts to shape public perception have become a defining feature of the modern information environment. Once rooted in military doctrine, such information operations (IOs) are now deployed by both state and non-state actors across civilian platforms to influence political discourse and erode institutional trust~\cite{dod_soie_2023,nato_information_threats_2024}. Contemporary policy frameworks emphasize that these activities extend beyond online disinformation and are not confined to digital channels, encompassing manipulative activity across the broader information environment~\cite{gao_information_environment_2022,eeas_fimi_2026,nato_information_threats_2024}.
Over the past decade, social media campaigns have evolved from relatively isolated spamming efforts into more sophisticated operations that exploit platform dynamics through impersonation, narrative fabrication, and cross-platform amplification~\cite{starbird2019disinformation, linvill2020troll, pacheco2021uncovering, cinus2025exposing}. Here, we focus on this platform-mediated subset, which provides a tractable and empirically observable setting for studying how such influence campaigns operate in practice.
A foundational distinction in the literature separates \emph{bots}---automated accounts executing scripted behaviors---from \emph{trolls}, human operators who deliberately manipulate interactions and narrative flow. While bots can spread disinformation at scale, human-operated accounts often play more central roles during high-impact events~\cite{gonzalez2021bots}. Both exploit attention dynamics by targeting influential users, amplifying low-credibility content, and triggering cascades that disproportionately boost visibility~\cite{shao2018spread}. These dynamics help explain why false content tends to spread more quickly and widely than factual information~\cite{vosoughi2018spread, juul2021comparing}, and why fake news during the 2016 U.S.~election was heavily concentrated among a small number of accounts~\cite{grinberg2019fake}. Recent advances in generative AI further lower the barrier to entry, enabling large-scale fabrication of persuasive content~\cite{spitale2023ai, goldstein2024persuasive}.

\noindent To address these evolving threats, detection research has historically focused on two primary signal types: \emph{content} (what is said) and \emph{network structure} (who interacts with whom). On platforms such as X/Twitter and Facebook, follower-followee relationships provide useful topological features. However, newer platforms---such as TikTok and Reddit---often lack explicit relational graphs, and even where such metadata exists, it is increasingly restricted due to platform policy and privacy concerns. As a result, public-facing activity logs---such as posts, comments, and timestamps---now constitute the primary empirical substrate for detection models. In parallel, malicious actors have adapted. Recent studies document persistent coordinated behavior across political and financial domains, from election interference to cryptocurrency manipulation~\cite{pacheco2020unveiling, pacheco2021uncovering, luceri2024unmasking}. These operations often exhibit synchronized timing, shared links, and repetitive behavioral sequences. As deplatforming pressures shift such actors to less-regulated environments~\cite{horta2023deplatforming}, the most stable observable signal is increasingly temporal: when and how users act. Content- and network-based detection methods are increasingly vulnerable to adversarial evasion, as actors modify surface features---such as encoded language and memetic structure---to bypass classification \cite{sayyadiharikandeh2020detection}. In response, researchers have shifted attention toward \emph{behavioral signals} as more resilient indicators of manipulation~\cite{cresci2020decade, luceri2024unmasking}, since action patterns are significantly harder to obfuscate than content. A growing literature focuses on behavioral coordination---recurrent patterns across accounts that include shared action sequences, synchronized posting, and temporal regularities. These signals tend to persist even when content is masked, and can be captured through temporal modeling frameworks~\cite{kong2023interval}. Such models often reveal underlying behavioral policies: stable, sequential patterns of interaction that are difficult to fabricate and resilient even in the absence of textual or relational cues. Together, this body of work suggests that user behavior---particularly in terms of timing and structure---offers more stable and generalizable detection signals than message content. This perspective aligns with a broader literature on influence operations spanning campaign structure, coordinated behavior, and account-level detection, which we discuss in detail in the \textit{Methods} section.

\noindent Building on this behavioral perspective, we investigate whether individual users can be reliably identified as malicious actors based on their inferred decision policies. Our focus is on binary classification: distinguishing troll accounts from ordinary users. Rather than uncovering coordination networks or content-based signals, we study whether sequential action patterns alone are sufficient for detection. Prior research has shown that trolls display distinctive behavioral signatures~\cite{luceri2020detecting}, and that such patterns can amplify strategic messaging in geopolitical contexts~\cite{geissler2023analyzing}. We extend these insights to Reddit, a pseudonymous platform where IO campaigns unfold not through follower graphs but through embedded participation in multiple subreddit communities.
We analyze the complete activity histories of 12,064 Reddit users---comprising over 38 million posts and comments collected between 2015 and 2018---including 99 accounts identified by Reddit's 2017 transparency report as linked to the Russian Internet Research Agency. We first document systematic differences in temporal rhythms and action preferences between troll and organic users. We model each user's interaction history as a sequence of \emph{states} and \emph{actions}, enabling the inference of personalized \emph{behavioral policies} that describe how users make platform decisions over time.
We formalize this process as a Markov Decision Process (MDP), where states encode recent engagement outcomes and actions represent platform functionalities such as initiating threads or replying to comments. A user's policy maps each state to a distribution over actions, yielding a compact representation of decision-making over time.
To infer user-level policies, we evaluate three approaches of increasing expressiveness: an \emph{empirical policy} based on observed state-action frequencies; a policy learned via \emph{Generative Adversarial Imitation Learning} (GAIL); and a maximum-entropy deep \emph{Inverse Reinforcement Learning} (IRL) method that recovers a reward function and derives a stochastic policy via soft value iteration. Each method produces a policy representation that we use as input to a supervised classifier. A policy-based classifier operates on these learned state-conditioned action distributions rather than on raw features, comparing them across users to distinguish behavioral patterns.

\noindent To benchmark behavior-based detection, we compare these policy representations to text embeddings derived from the same user traces. This parallel design enables a controlled comparison of behavioral and content-based signals across matched cohorts and observation windows. We evaluate all four representations on classification accuracy, early detection from limited activity, and robustness to textual corruption or evasion.
Finally, we examine within-group heterogeneity using clustering analyses and investigate failure modes, such as hijacked accounts or trolls that closely mimic organic behavior. Together, our findings show that behavioral policies offer stable signals for identifying IO actors on Reddit, and provide a resilient foundation for detection---particularly in settings where content is obfuscated or metadata is unavailable.

\section*{Results}\label{sec:results}

We analyze behavioral differences between troll and organic users across two levels of abstraction: raw activity patterns and inferred decision policies. We first compare descriptive statistics of user behavior---timing, frequency, and type of interaction---revealing systematic differences between trolls and organic users in how they engage with the platform. We then infer individualized policies that summarize each user's strategic behavior as a sequence of state-dependent actions. These policies form the basis for classification and robustness analysis in subsequent sections.

\subsection*{Troll user activity patterns}
\label{sec:activity-patterns}

A first step toward modeling sequential decision processes is to identify behavioral regularities that distinguish trolls from ordinary users.

\begin{figure}[!htbp] 
    \centering
    \includegraphics[width=0.9\textwidth]{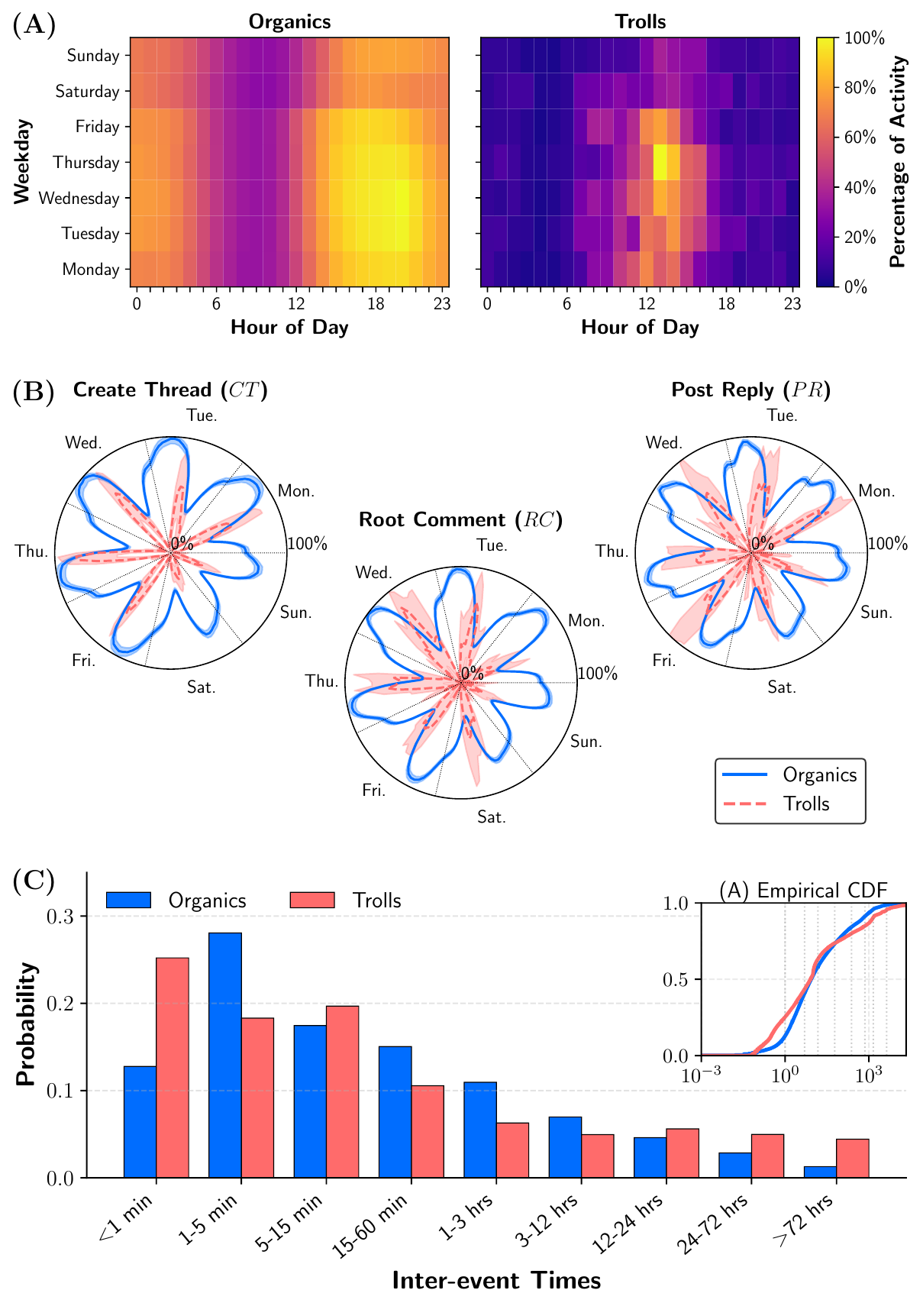}
    \caption{\textbf{Activity patterns of organic users and trolls (Reddit dataset).} 
    \textbf{(A)} Heatmaps of hourly activity (UTC). Trolls peak in U.S.~morning hours; organics peak in U.S.~afternoon.
    \textbf{(B)} Radial plots of weekly timing for three actions (create thread, root comment, post reply). Trolls exhibit sharper, synchronized morning peaks across all actions, consistent with operational scheduling.
    \textbf{(C)} Inter-event time distributions. Trolls show a heavy-tailed, bursty pattern (fast actions followed by long dormancy), whereas organics follow a Weibull-like distribution.}
    \label{fig:activity-patterns}
\end{figure}

\noindent Figure~\ref{fig:activity-patterns} (A) compares the temporal distribution of activity for organics (left) and trolls (right), aggregated by weekday and hour (UTC). Each panel is normalized to its own maximum (0--100\%) to highlight relative hot spots. Organic activity reflects the rhythms of Reddit's predominantly U.S.-based user base: engagement rises around 12:00~UTC (08:00~EST) and declines after 04:00~UTC (00:00~EST), with weekends following a similar structure at lower intensity. Troll activity, in contrast, exhibits a pronounced peak between 12:00 and 14:00~UTC---corresponding to early morning hours in the U.S.---suggesting a strategic effort to shape the early flow of daily discussions. This observation aligns with prior work documenting operational schedules embedded in the temporal signatures of state-linked campaigns~\cite{linvill2020troll, zannettou2019web}.

\noindent These patterns indicate that trolls are not merely participating in discussions but are timing their activity to align with periods of maximum exposure to U.S. audiences. Rather than maintaining a constant presence, these state-linked influence accounts concentrate their actions within specific time windows, consistent with coordinated, goal-driven deployment. Differences also arise in the mix of actions they perform. As shown in Figure~\ref{fig:activity-patterns} (B), both organic users and troll accounts display weekly activity rhythms across key behaviors---creating new threads, posting top-level comments, and replying. Each panel presents a radial view of one action type, where the full circle represents a week (Monday-Sunday), the angular position encodes the time of day, and the radius reflects activity normalized within each group, such that the outer ring denotes peak activity, with shaded bands indicating confidence intervals. However, trolls exhibit markedly sharper and more synchronized patterns, with activity clustering during U.S.~morning hours across all categories. This structured alignment is consistent with coordinated operational schedules typical of adversarial information operations.

\noindent Beyond daily and weekly rhythms, inter-event timing---the distribution of time intervals between consecutive user actions---captures fine-grained temporal dynamics that further differentiate trolls from organics. As shown in Figure~\ref{fig:activity-patterns} (C), the two groups exhibit significantly different timing profiles ($p<0.001$, two-sample Anderson--Darling test), suggesting distinct patterns of temporal regularity and engagement structure.

\noindent For organics, the distribution closely resembles a Weibull form: short intervals are rare, reflecting the perceptual and cognitive latency required to process and respond to content. The distribution then rises and decays gradually, consistent with memory effects in online attention~\cite{rizoiu2017expecting}. Troll activity, by contrast, exhibits a heavy-tailed power-law distribution. A substantial share of actions occur within seconds of a prior post, indicating minimal response latency and potentially reflecting pre-scripted or automated posting. The long tail reflects extended inactive periods, which may arise from accounts entering periods of operational dormancy and later being selectively reactivated, or from intentional re-engagement with specific content as part of a long-term agenda-setting strategy.

\noindent Trolls are significantly more likely to post again within one minute ($25.2\%$ vs.~$12.8\%$; Fisher's exact test, $p<10^{-6}$) and to remain inactive for over 72 hours ($4.4\%$ vs.~$1.3\%$; Fisher's exact test, $p<10^{-6}$). This alternation between high-frequency bursts and prolonged dormancy is rarely observed among organics. It is consistent with coordinated operations that rely on tightly timed posting windows interspersed with inactivity to maintain synchronization and limit detection~\cite{cinelli2022coordinated, zannettou2019disinformation, bellutta2023investigating} (see \textit{Supplementary Information} for further dataset details).

\subsection*{Activity-based detection of malicious troll accounts}
\label{subsec:detection}

\begin{figure}[!htbp] 
    \centering
    \includegraphics[width=0.85\textwidth]{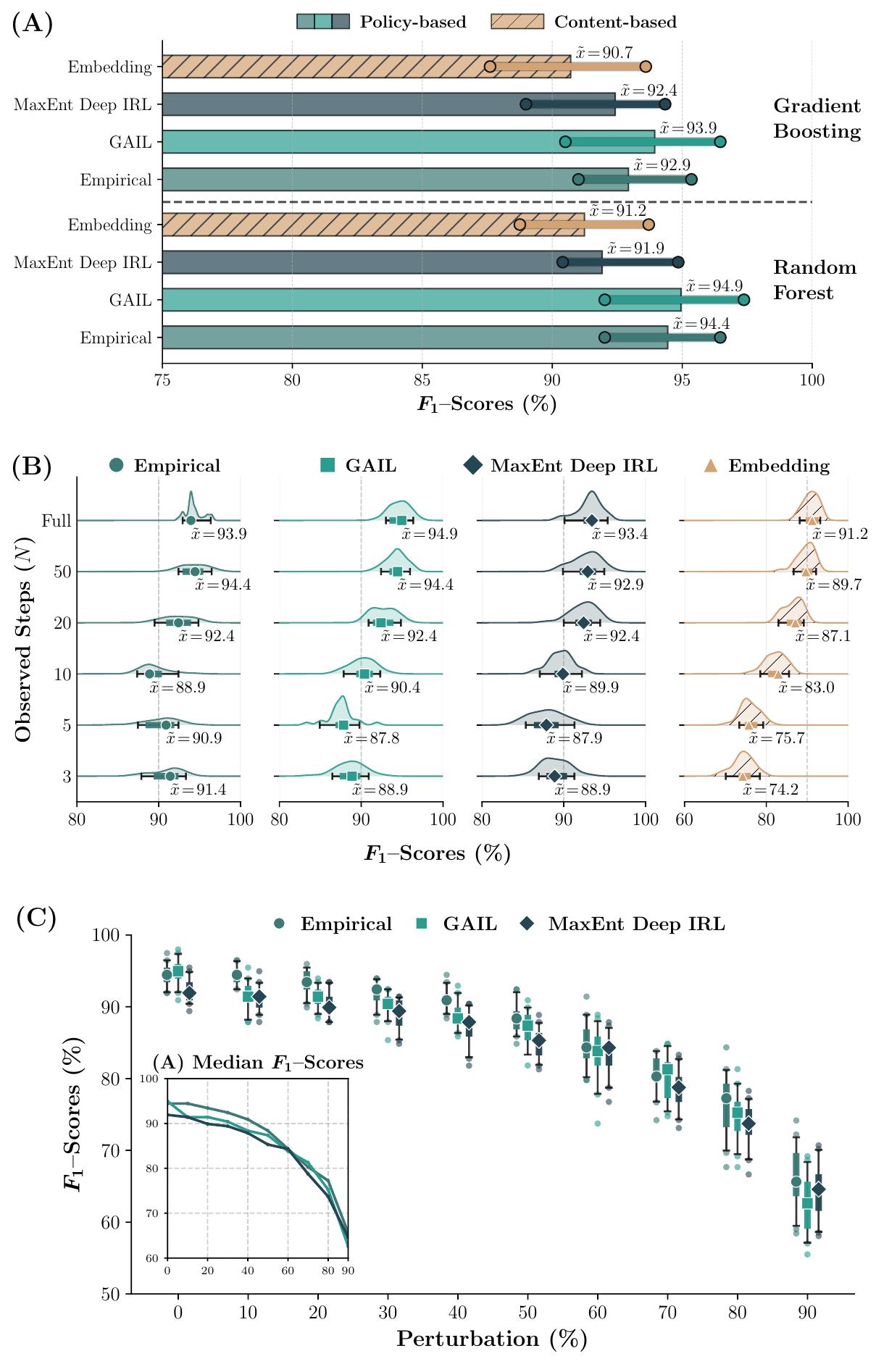}
    \caption{\textbf{Policy-based vs.~content-based troll detection.} \textbf{(A)} Classification performance (median macro-$F_1$, 5th--95th percentiles). Policy-based methods (Empirical, IRL, GAIL) consistently outperform the text Embedding baseline across classifiers. GAIL achieves the highest performance (94.9\%). \textit{Note:} Organic trajectories were truncated to match troll lengths. \textbf{(B)} Performance by trajectory length ($n$ state-action pairs). Policy-based methods outperform content-based at all $n$, with the largest margins at small $n$ (\textit{e.g.}, $n{=}3$: Empirical 91.4\% vs. Embedding 74.2\%). \textbf{(C)} Robustness to perturbed trajectories. Median macro-$F_1$ as perturbation level $p$ increases. All methods degrade with noise; Empirical maintains the highest median across most $p$.}
    \label{fig:policy-vs-content}
\end{figure}

\noindent\textit{Policy-based vs content-based classification.}
Figure~\ref{fig:policy-vs-content} (A) reports macro-$F_1$ scores for all methods (empirical policy, IRL, and GAIL), evaluated under stratified $k$-fold cross-validation, which preserves the troll/organic imbalance across folds (see \textit{Methods}; complete experimental details and additional results are provided in Supplementary Information).

\noindent Policy-based representations consistently outperform content-based baselines, regardless of the underlying classifier---eXtreme Gradient Boosting (XGBoost) or Random Forest (RF). Overall, RF, as an ensemble method, achieves higher median performance and narrower 90\% confidence intervals than XGBoost for most policy estimators, with the exception of maximum-entropy deep IRL.

\noindent GAIL attains the highest performance, with a median macro-$F_1$ of 94.9\% (5th--95th percentiles: 92.0-97.4). All policy-based approaches exceed 90\% median $F_1$, indicating that labeled troll accounts exhibit distinct and reproducible behavioral patterns across subreddits. The behavioral homogeneity observed within the troll cohort is consistent with agents following scripted or tightly constrained interaction strategies---markedly different from the more diverse behavior of organic users \cite{geissler2023analyzing, pote2025coordinated}.

\noindent\textit{Detection from short activity traces.}
Online content has a brief attention lifecycle: most impressions occur within hours of posting \cite{pfeffer2023halflife}, after which diffusion declines rapidly. 
As a result, Information Operations often inflict the most harm shortly after deployment, making early detection of malicious actors essential to reducing the overall impact of attacks.

\noindent However, identifying troll accounts early---before they have left a long behavioral trace---poses a challenge for detection systems due to data scarcity. 
To evaluate performance under such constraints, we train classifiers on only the first $n$ user actions (submissions or comments), where each action is represented as a state-action pair in our model, varying $n$ from 3 to the full activity history (Fig.~\ref{fig:policy-vs-content} (B)).
Policy-based methods consistently outperform content-based baselines across all $n$, with the largest margins observed at small $n$, demonstrating that early behavioral signals are highly informative.

\noindent Even trajectories with just three state-action pairs yield strong predictive performance: the Empirical policy achieves a macro-$F_1$ of 91.4\% (5th--95th percentiles: 87.9-93.3), outperforming GAIL at 88.9\% and IRL at 88.9\%, both with narrow confidence intervals. All policy-based methods substantially exceed the content-based Embedding baseline (74.2\%, 5th--95th percentiles: 70.1-78.4), with non-overlapping intervals. A qualitative examination of early activity patterns indicates that trolls frequently begin by initiating multiple new threads in quick succession, whereas organic users more commonly start by replying within existing discussions. Creating a thread results in immediate visibility, and new users---who are still adapting to platform norms---tend to avoid this action initially. The state-action representation captures these early strategic choices directly, enabling strong separation between trolls and organics even under minimal observation.

\noindent Performance stabilizes quickly: by $n = 10$-20, all policy-based methods reach macro-$F_1$ scores of 92.4\% or higher, while the content-based method lags behind at 87.1\% (5th--95th percentiles: 83.0-89.2). With longer histories ($n{=}50$), GAIL and Empirical reach 94.4\% (5th--95th percentiles: 92.4-96.0 and 92.4-96.5), IRL achieves 92.9\% (5th--95th percentiles: 89.9-94.9), and Embedding improves to 89.7\% (5th--95th percentiles: 86.7-92.2). Notably, confidence intervals remain non-overlapping: Embedding's upper bound (92.2\%) does not intersect GAIL's lower bound (92.4\%).

\noindent When using the full trajectory, performance peaks at 94.9\% for GAIL, 93.9\% for Empirical, and 93.4\% for IRL, all outperforming Embedding (91.2\%). Across all observation windows, policy-based models provide higher accuracy and lower variance---delivering early, stable detection from as few as three user actions.

\noindent\textit{Robustness to observational noise.} 
Social media data is inherently noisy---even without malicious intent. Posts may be logged out of order, timestamps can drift, and user traces may be incomplete. In adversarial settings such as Information Operations, this noise is further amplified by actors who actively seek to obfuscate their behavior. Detection systems must therefore remain effective even when behavioral sequences are partially corrupted. To assess robustness, we inject noise by replacing the action at a fraction $p$ of timesteps with a uniformly sampled action from $\mathcal{A}$ and resampling the subsequent state from the set of legal next states.
Figure~\ref{fig:policy-vs-content} (C) shows that policy-based models degrade gradually as \(p\) increases. 
All maintain macro-$F_1$ scores above 80\% up to \(p{=}50\%\), with Empirical features showing better stability than IRL or GAIL. 
For example, at \(p{=}10\%\), Empirical drops just 1.0 point (from 94.4\% to 93.4\%), while GAIL and IRL fall to 91.4\% and 89.9\%, respectively. By \(p{=}40\%\), these methods reach 90.9\% (Empirical), 88.4\% (GAIL), and 87.9\% (IRL), with confidence intervals gradually widening. At extreme corruption levels ($p{\ge}70\%$), performance converges across all methods into the low 60s to low 80s, with occasional reversals in relative ranking. 
Nevertheless, degradation remains smooth, not brittle, suggesting that policy-based representations retain meaningful structure even under significant distortion. 
Overall, these results confirm that policy features are both predictive and robust: they degrade gracefully under noise and retain strong discrimination power in the early, most consequential phases of user activity---critical qualities for real-world detection in adversarial and imperfect environments.

\subsection*{Behavioral heterogeneity among trolls}

Troll behavior diverges systematically from that of organic users. Earlier sections documented distinct temporal activity patterns and sequential decision strategies that enable highly accurate classification based on inferred policies. \textit{Are all trolls behaviorally alike? And how do their policies differ, even in simplified representations such as action distributions?} To investigate this, we qualitatively analyze the policies of all 99 troll accounts in our dataset.

\noindent\textit{Troll clusters.}
Although group-level summaries highlight aggregate differences, individual-level behavior may vary substantially. In the case of trolls---whose actions may reflect partially scripted routines or coordinated templates---distinct patterns still emerge. Applying $k$-means clustering to the GAIL-inferred policies of the 99 troll accounts yields three consistent behavioral subgroups. To assess the distinctiveness of each cluster, we compare its members to a matched sample of 99 organic users selected for similar overall activity levels. 

\begin{figure}[!htbp]
    \centering
    \includegraphics[width=0.78\textwidth]{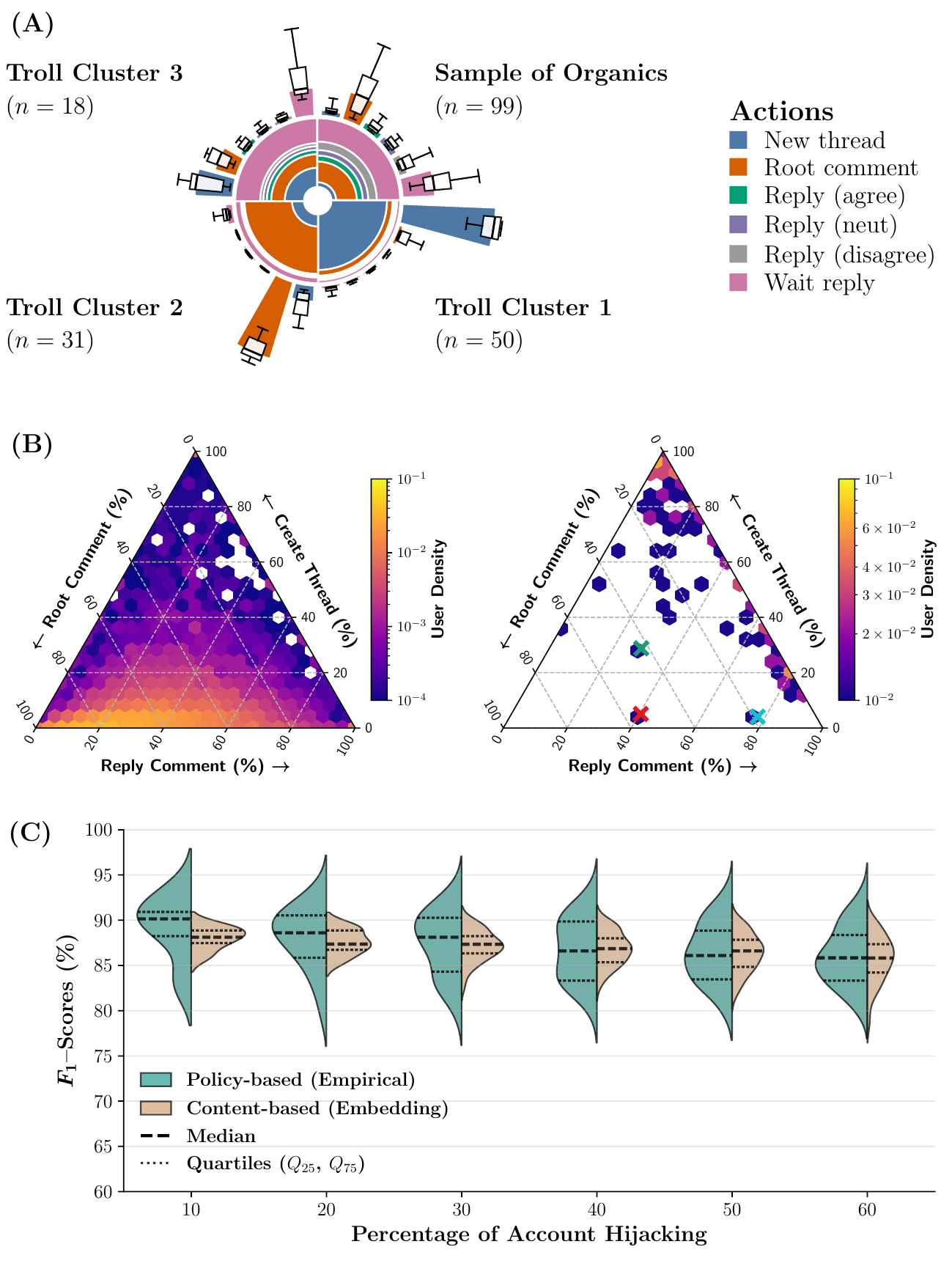}
    \caption{\textbf{Analysis of troll behavior and adversarial robustness.}
    \textbf{(A)} Clustered behavior of three troll clusters vs organics. Inner circle shows the mean distribution of actions by cluster. Outer bars show action distributions for all users in each cluster, with boxplots indicating 5th--95th percentiles. Clusters 1 and 2 specialize in thread creation (84.6\%) and root comments (72.5\%), respectively. Cluster 3 resembles organics but with elevated thread creation (34.4\% vs. 4.9\%). Replies are underrepresented across all troll clusters.
    \textbf{(B)} User distribution of action profiles. Organics (left) display a more balanced profile; trolls (right) concentrate on thread creation and root comments. Three consistently misclassified trolls (\xmark[black]) exhibit organic-like, reply-dominated profiles.
    \textbf{(C)} Account hijacking. Policy-based vs. content-based performance by hijacking severity ($\eta$ = fraction organic activity). Policy-based outperforms at early hijacking ($\eta=10\%$: 90.1\% vs. 88.1\%); performance converges at $\eta=50\%$ with higher policy variance.}
    \label{fig:troll-behavior}
\end{figure}

\noindent Figure~\ref{fig:troll-behavior} (A) compares the action composition of organic users with three troll policy clusters, with each quarter representing one group; each user's policy is expressed as normalized frequencies across six action categories. The inner ring shows the mean distribution across these actions, whereas the outer bars summarize how these action shares vary across users, indicating the median, interquartile range and 5th-95th percentiles. Two clusters exhibit highly concentrated behavior. Cluster 1 (50 accounts) overwhelmingly favors thread creation (median: 84.6\%), while Cluster 2 (31 accounts) prioritizes root comments (median: 72.5\%). 
In both, reply actions are nearly absent, indicating a strategy centered on content initiation and visibility rather than dialogue. 
These findings align with prior studies \cite{geissler2023analyzing, pote2025coordinated} documenting how state-linked actors amplify narratives through thread injection and resharing while avoiding interactive engagement. Cluster 3 (18 accounts), by contrast, resembles organic users more closely. For example, the probability mass assigned to the wait-reply action is 29.3\%, compared to 25.3\% for organics. Nevertheless, differences persist: thread creation remains elevated (34.4\% vs. 4.9\%), and reply actions remain underrepresented. A consistent trait across all troll clusters is the scarcity of replies. On Reddit, the first reply to a thread is referred to as a ``root comment''---a platform-specific convention that differs from others such as X (formerly Twitter), where such distinctions are not formally labeled.
This broader pattern points to a unifying behavioral strategy oriented toward broadcasting narratives rather than participating in multi-turn interactions.

\noindent While most troll accounts were successfully identified by our models, three---anonymized as A (\xmark[TrollRed]), B (\xmark[TrollTeal]), and C (\xmark[TrollCyan])---were consistently misclassified as organic. In Figure~\ref{fig:troll-behavior} (B), these accounts appear as colored \xmark[black] markers matching their labels. Their GAIL-inferred policies differ markedly from the broader troll population. All three assign low probability to thread creation (A: 5.1\%, B: 28.8\%, C: 4.1\%) and instead focus on replying. Account C, for instance, mirrors organic user behavior with a distribution of 4.1\% thread creation, 18.4\% root comments, and 77.5\% replies.

\noindent A qualitative review of their content reveals no clear markers of state-sponsored influence, suggesting that these accounts were either falsely flagged or identified based on platform-level metadata not available to us (\textit{e.g.}, IP addresses, geolocation, device identifiers). 
One plausible hypothesis is that such accounts function as ``personal'' or ``cover'' profiles for authentic engagement, distinct from others used in coordinated information operations---a separation strategy frequently observed in adversarial campaigns.

\noindent\textit{Robustness against detection evasion.}  
All detection systems are vulnerable to adversarial evasion. Unlike content-based classifiers---which treat user activity as an unordered collection of observations---our behavioral framework depends on the temporal structure and order of state-action sequences. This reliance introduces susceptibility to obfuscation strategies that aim to mimic organic behavior.

\noindent One such strategy is \emph{account hijacking}, in which malicious actors gain control of legitimate user profiles and embed targeted actions within otherwise benign activity histories. Another tactic, known as \emph{ephemeral astroturfing}---short-lived, coordinated bursts of activity using disposable or repurposed accounts to simulate grassroots support---produces transient behavioral signals that can evade detectors dependent on persistent lexical or network structure \cite{elmas2021ephemeral}. While these tactics complicate detection, prior work suggests that behavioral profiling can improve identification of compromised accounts \cite{ruan2016profiling}.

\noindent To assess resilience under this threat model, we simulate hijack scenarios by generating synthetic trajectories using the policies of trolls and organic users.
A hijacked user's trajectory is constructed by generating the initial proportion $\eta$ of actions using an organic user's policy, with the remaining actions generated under a troll's policy.
Non-hijacked users are generated entirely using an organic policy.
We vary $\eta$ between 10\% and 60\% to control the fraction of organic activity in each trajectory, thereby simulating how early or late in the account lifespan the hijacking occurs; lower values of $\eta$ correspond to earlier hijackings and shorter organic histories.
Policies are then re-inferred from the generated trajectories using the empirical policy estimator, which we found to be the most stable method under perturbations.

\noindent Results are shown in Figure~\ref{fig:troll-behavior} (C), comparing the classification performance of the empirical policy against content embeddings.
The policy-based method performs better in the evasion scenario when the account is hijacked early in its lifespan and therefore contains less organic activity.
At 10\% organic activity, the empirical-policy classifier achieves a median $F_1$ of 90.1\% (95\% CI: 86.9--91.3; 5th--95th percentiles: 83.0--93.0), while the embedding-based classifier attains 88.1\% (95\% CI: 86.0--89.4; 5th--95th percentiles: 86.0--89.4), indicating slightly stronger performance for the policy-based approach.
At 50\% organic activity, the policy-based approach maintains comparable median performance but exhibits substantially higher variance as the proportion of organic activity increases (Policy: median $F_1$ 86.1\%, 95\% CI: 85.5--87.3; Content: median $F_1$ 86.6\%, 95\% CI: 85.8--86.9).
As expected, classification performance deteriorates as a larger share of the trajectory is replaced with organic behavior, reducing the presence of discriminative policy signals.
This, in turn, increases output variance, as fewer informative samples amplify the impact of noise and reduce classifier confidence.

\section*{Discussion}
\label{sec:discussion}

The online information ecosystem is undergoing rapid transformation: social platforms optimized for rapid dissemination and community formation now mediate a large share of public attention, while established newspapers, television news and public broadcasters have declined in relative influence. 
Empirical studies show that false and inflammatory content spreads quickly and widely on these platforms \cite{del2016spreading, lazer2018science}, and that coordination and timing can significantly amplify both reach and persuasive impact \cite{tardelli2024temporal, bail2020assessing}. These effects are not only structural but also cognitive: repeated exposure increases perceived accuracy---a phenomenon known as the \emph{illusory truth effect} \cite{hasher1977frequency}---while conformity pressure can drive individuals to adopt prevailing group opinions, even when inaccurate \cite{pennycook2018prior, lorenz2011social}. As a result, well-timed and highly visible messages disproportionately shape public beliefs.
At the same time, shrinking access to platform data and rapid advances in generative text reduce the reliability of content- and graph-based defenses, leaving detection systems vulnerable as malicious actors adapt. This combination of limited data access and shifting tactics also undermines reproducibility and external validation of detection methods.

\noindent Our results show that behavioral dynamics---operationalized as state-action trajectories and recovered policies---constitute a complementary and robust signal for identifying malicious actors. 
Using a dataset of 12,064 Reddit users (including 99 accounts linked to the Russian Internet Research Agency) and more than 38 million activity steps, we report three main empirical findings. First, troll accounts exhibit distinctive temporal rhythms and inter-event dynamics (posting in bursts with long dormancy) that diverge markedly from organic users. 
Second, policy-based representations inferred via empirical state-action frequencies, Generative Adversarial Imitation Learning (GAIL) and maximum-entropy deep IRL produce user-level policies that outperform a state-of-the-art text-embedding baseline (GAIL median macro-$F_1$ 94.9\% vs text-embedding 91.2\%), achieve strong performance from very short traces (empirical policy 91.4\% macro-$F_1$ with three state-action pairs), and degrade gradually under random corruption and simulated hijacking. A comparison with closely related prior work is provided in the \textit{Supplementary Information}, which contextualizes our results within the broader literature on account-level detection. 
Third, behavioral heterogeneity exists within the troll cohort: clustering reveals multiple operational styles (for example, thread-creation versus root-comment strategies) and a small set of evader accounts whose policies closely resemble organic users, highlighting specific failure modes.

\noindent These findings have both theoretical and practical implications. On the practical side, behavior-based features allow for earlier intervention, which is critical given the short attention half-life of online posts \cite{pfeffer2023halflife, schneider2023effectiveness}. For instance, on platforms such as Twitter, the attention half-life of a post is often as short as 24 minutes. Moreover, behavior-based approaches are more robust than content-only methods when faced with evasion tactics such as injected benign activity or account hijacking \cite{thomas2014consequences, bursztein2014handcrafted, ruan2016profiling}. The graceful degradation we observe indicates operational utility even under noisy or partially observed conditions. Theoretically, modeling activity as a sequential decision process emphasizes temporal structure as an axis of manipulation and provides a platform-agnostic framework for comparing tactics on venues without explicit follower graphs; cross-platform transfer is plausible but will depend on event granularity and logging conventions \cite{luceri2020detecting, geissler2023analyzing}. Together, these results underscore the value of behavioral policies as a complement to content- and network-based systems, motivating future hybrid pipelines that combine behavior, text, and vetted platform metadata for improved real-world detection.

\noindent Important limitations remain. This study focuses on Reddit, a platform structurally distinct from the X/Twitter-centric majority of the IO detection literature. We analyze Reddit's 2017 transparency report (published April 2018), which provides the only longitudinal, user-level IO labels available on Reddit; these labels reflect the platform's internal detection heuristics and may be incomplete or skewed toward tactics easier to identify automatically. A small subset of accounts was consistently misclassified, suggesting that malicious actors can operate with behavioral footprints closely mimicking organic users---either by design or to serve distinct operational roles (\textit{e.g.}, ``cover'' versus ``work'' profiles). Because the model is event-indexed rather than duration-aware, inter-event timing is not represented explicitly in the state space; timing differences enter only through action sequences and the frequency of $WR$ selections. While sufficient for classification, duration-aware (\textit{e.g.}, semi-Markov) extensions may be useful when explicit waiting-time structure is central. Applying this framework to other platforms will require platform-specific state representations and transfer-learning protocols; progress depends on standardized event schemas and cross-platform benchmarks.

\noindent These caveats suggest three focused priorities. First, treat behavioral features as a complement to text: our experiments show that simple state-action embeddings reach high accuracy from only a few actions and reduce variance relative to text alone, so practical systems should combine short-window behavioral summaries with text embeddings and route low-confidence cases to analyst review rather than automatic enforcement. 
Second, evaluate and strengthen models against the concrete failure modes we identified: the random-swap and simulated-hijack tests, together with the misclassified accounts, expose strategies for targeted evasion, and red-teaming protocols that inject benign activity, swap intra-account actions, and simulate ephemeral astroturfing should be incorporated into model selection and calibration. 
Third, prioritize cross-platform transferability and privacy-preserving evaluation; extending the framework requires adapting state representations to platform structure while minimizing raw-data exposure (\textit{e.g.}, via federated training on hashed state-action counts or releasing differentially private summaries).

\noindent In summary, modeling how actors behave, in addition to what they post, provides a complementary signal for detection. Policy-based representations are accurate and data-efficient in this setting; when combined with content, metadata, and human oversight, they strengthen practical defenses against evolving information campaigns.

\section*{Methods}
\label{sec:methods}

\subsection*{Related Work}
Prior work on online influence operations is organized around three core problems: characterizing campaigns and actor roles, identifying coordinated groups, and detecting individual accounts, with a related security literature on compromised accounts. These distinctions do not align directly with terminology such as bots, coordinated inauthentic behavior, and influence operations. We study the empirical problem of distinguishing IRA-linked Reddit troll accounts from organic users using each account's public state-action trajectory.
\smallskip 

\noindent\textit{Campaign structure and role differentiation.} A first line of work studies influence operations at the level of campaigns rather than individual accounts. Starbird \textit{et al.}\ argue that disinformation is collaborative work: false or misleading narratives are sustained by distributed participants who play different roles in producing, amplifying, and legitimizing content \cite{starbird2019disinformation}. Focusing on Russian Internet Research Agency activity on Twitter, Linvill and Warren show that troll accounts adopted specialized personas with distinct behavioral roles rather than behaving as a uniform class of malicious users \cite{linvill2020troll}. Atanasov \textit{et al.}\ formalize this heterogeneity by predicting troll roles from social and textual traces \cite{atanasov2019predicting}. Zannettou \textit{et al.}\ characterize state-sponsored troll behavior on Twitter and trace how that activity propagates into other web communities, showing that operational effects are not confined to a single platform \cite{zannettou2019disinformation}. Badawy \textit{et al.}\ analyze the 2016 IRA campaign through retweet networks, ideological structure, and bot prevalence among accounts interacting with troll content \cite{badawy2019characterizing}. Zannettou \textit{et al.}\ later show that these campaigns are also multimodal, using images and memes strategically around real-world events \cite{zannettou2020characterizing}. Bail \textit{et al.}\ shift the focus from campaign design to audience response by examining exposure to IRA accounts and reporting limited short-run attitudinal effects in late 2017 \cite{bail2020assessing}. This literature shows that influence operations are organized, role-differentiated, and adaptive. However, it does not address account-level detection based solely on a single user's public trajectory.
\smallskip 

\noindent\textit{Detecting coordinated groups.} A second line of work aims to recover groups of accounts that act together. Here the key signal is not necessarily message meaning, but shared behavior. Giglietto \textit{et al.}\ operationalize coordinated link sharing on Facebook by identifying repeated co-linking of the same URLs within narrow time windows \cite{giglietto2020coordinated}. In follow-up work on the 2018 and 2019 Italian elections, they show that this strategy transfers to concrete electoral cases in which repeated URL sharing reveals organized amplification around political events \cite{giglietto2020takes}. Their later workflow paper turns this insight into an operational pipeline for detecting, monitoring, and updating coordinated account lists over time \cite{giglietto2023workflow}. Pacheco \textit{et al.}\ use shared actions to uncover coordinated groups behind White Helmets disinformation \cite{pacheco2020unveiling}, and then generalize this idea by constructing coordination networks from arbitrary shared behavioral traces across several case studies \cite{pacheco2021uncovering}. Sharma \textit{et al.}\ argue that coordinated accounts can be identified through hidden influence relations and group behavior even when content is noisy or strategically varied \cite{sharma2021identifying}. Nizzoli \textit{et al.}\ apply related ideas to the 2019 UK general election, showing that synchronized manipulation can be measured in an election-specific setting \cite{nizzoli2021coordinated}. Jakesch \textit{et al.}\ extend the problem across platforms by tracing how an external organization manipulated Twitter trends during the Indian general election \cite{jakesch2021trend}. Cinelli \textit{et al.}\ add a downstream perspective by connecting coordinated inauthentic behavior to patterns of information diffusion on Twitter \cite{cinelli2022coordinated}.

\noindent Recent work sharpens the conceptual boundaries of this literature. Polychronis and Kogan argue that state-sponsored operations should not be collapsed into the generic social-bot paradigm because such campaigns often rely on division of labor, collaboration, and mixed human--automated participation \cite{polychronis2023working}. Tardelli \textit{et al.}\ show that coordinated communities differ in temporal stability and participation archetypes \cite{tardelli2024temporal}; in later work, they show that coordination is multifaceted and not reducible to a single feature family \cite{tardelli2024multifaceted}. Luceri \textit{et al.}\ use fused behavioral similarity networks to identify information-operation drivers at scale \cite{luceri2024unmasking}. Cinus \textit{et al.} extend coordination analysis across platforms, showing evidence of coordinated inauthentic activity across X, Facebook, and Telegram in the run-up to the 2024 U.S. election \cite{cinus2025exposing}.

\noindent For our purposes, the central lesson is that manipulation often becomes visible through temporal and behavioral regularities even when content varies. At the same time, these approaches are predominantly group-centric: they depend on cross-account similarity, shared links, interaction networks, or synchronized activity. Our setting is narrower and complementary: we ask whether reliable detection is possible without explicit cross-account structure, by classifying each account from its own public trajectory.
\smallskip 

\noindent\textit{Individual-account detection.} A third line of work treats malicious-account detection as an account-level classification problem. Early work such as Truthy showed that large-scale analysis of meme diffusion could surface political astroturfing in microblog streams \cite{ratkiewicz2011truthy}. The social-bot literature then defined many of the standard baseline families for this task. Ferrara \textit{et al.}\ synthesize the rise of social bots and organize the field around content, network, temporal, and profile-based features \cite{ferrara2016rise}. BotOrNot operationalizes bot scoring through a rich inventory of linguistic, sentiment, network, metadata, and timing cues \cite{davis2016botornot}. Varol \textit{et al.}\ extend this line by estimating bot prevalence at scale and examining human--bot interaction patterns \cite{varol2017online}. Cresci \textit{et al.}\ show that social spambots increasingly mimic ordinary users well enough to evade simple automation heuristics, motivating behavioral approaches such as social fingerprinting based on action sequences \cite{cresci2017paradigm,cresci2017social}. Sayyadiharikandeh \textit{et al.}\ address generalization to unseen bot types through ensembles of specialized classifiers \cite{sayyadiharikandeh2020detection}. Cresci's later review documents that evaluation settings, label definitions, and dataset construction vary so widely across studies that direct cross-study comparison of reported metrics is rarely meaningful \cite{cresci2020decade}.

\noindent These studies establish the main baseline approaches for account-level detection, but remain primarily focused on generic automation rather than state-sponsored troll accounts.

\smallskip

\noindent\textit{Behavior-first and Reddit-specific account detection.}
The closest comparators are papers that target influence-operation accounts more directly. Alizadeh \textit{et al.}\ show that content-based features can predict state-sponsored accounts on both Twitter and Reddit with high accuracy (macro-$F_1$ ranging from $0.74$ to $0.99$ across campaigns and platforms) \cite{alizadeh2020content}. Luceri \textit{et al.}\ move toward behavioral modeling by treating troll activity as a sequential decision process learned via inverse reinforcement learning, achieving an AUC of 89.1\% on IRA-linked Twitter trolls \cite{luceri2020detecting}. Ezzeddine \textit{et al.}\ likewise emphasize behavioral sequences and feedback signals, arguing that such signals become more valuable as synthetic text becomes easier to imitate \cite{ezzeddine2023exposing}. Luceri \textit{et al.}\ later show that large language models can support detection when metadata and network structure are serialized as inputs \cite{luceri2024leveraging}. On Reddit, Saeed \textit{et al.}\ introduce TrollMagnifier, which identifies Russian-sponsored troll accounts (per Reddit's 2017 transparency report, published April 2018) by expanding a seed set of known trolls through cross-account behavioral similarity, achieving $F_1$ of 97.8\% \cite{saeed2022}. Yuan \textit{et al.}\ develop a public state--action encoding of Reddit user trajectories and use IRL-derived policies to study behavioral homophily on Reddit \cite{yuan2025behavioral}.
\smallskip

\noindent Our study is closest to this latter group but differs in several respects. Relative to Yuan \textit{et al.}~\cite{yuan2025behavioral}, we use the state--action encoding for troll-account detection rather than for community-level analysis. Relative to TrollMagnifier \cite{saeed2022}, we do not rely on seed expansion or inter-account similarity, but treat each account as the primary unit of analysis. Relative to Luceri \textit{et al.} \cite{luceri2020detecting}, we study Reddit rather than Twitter and compare multiple policy-learning formulations on a shared benchmark. More broadly, our approach shifts the focus from static feature representations toward modeling user behavior as a dynamic decision process over time.
\smallskip

\noindent\textit{Diffusion, exposure, and account integrity.} Two adjacent literatures clarify why this task matters. Studies of diffusion and exposure show that malicious accounts can disproportionately shape information flows. Shao \textit{et al.}\ show that bots amplify low-credibility content, especially early in diffusion \cite{shao2018spread}. Stella \textit{et al.}\ show that bots increase exposure to negative and inflammatory material \cite{stella2018bots}. Vosoughi \textit{et al.}\ document that false news spreads faster and farther than truthful news online \cite{vosoughi2018spread}. Grinberg \textit{et al.}\ show that fake-news exposure during the 2016 U.S.\ election was highly concentrated among a small group of users \cite{grinberg2019fake}. 

\noindent A parallel security literature studies malicious behavior emerging from compromised or hijacked accounts. Thomas \textit{et al.}\ characterize account hijacking at scale \cite{thomas2014consequences}, while Bursztein \textit{et al.}\ document manual hijacking workflows \cite{bursztein2014handcrafted}. Ruan \textit{et al.}\ and Egele \textit{et al.}\ show that deviations from historical behavioral profiles provide robust signals of compromise \cite{ruan2016profiling,egele2015towards}. These findings motivate not only accurate detection, but also robustness to mixed, perturbed, or partially hijacked traces.
\smallskip

\noindent This literature points to a precise gap. Campaign-level studies explain how influence operations are organized, coordination studies recover sets of accounts that act together, and many account-level detectors rely on text, metadata, or cross-account structure. The closest prior work begins to use behavioral modeling for troll detection, but typically still relies on platform-specific features or inter-account signals. What remains less well understood is whether a \emph{single account's public state--action trajectory}, without explicit coordination features, is sufficient for reliable troll-account classification on Reddit. The present study examines this question by comparing empirical policies with those learned via inverse reinforcement learning and generative adversarial imitation learning on a matched Reddit benchmark, and by evaluating their performance under short-trace and perturbation settings.

\subsection*{Dataset}

Our dataset consists of activity from troll and organic user accounts collected via the Pushshift Reddit API \cite{baumgartner2020pushshift}, covering the period from the beginning of 2015 through the end of 2018.
Troll accounts were identified using Reddit's 2017 transparency report (published April 2018), which listed 944 accounts as suspicious due to links to the Russian Internet Research Agency, an organization established for the deliberate purpose of conducting information operations. 
Based on these account names, we retrieved both \emph{active behavior} (content generation, \textit{i.e.}, submissions) and \emph{passive behavior} (first-order responses, such as receiving comments on a submission). 
The precise definitions of the action and state spaces, including these action types, are provided in the following section.

\noindent Of the 944 accounts named in the report, 797 had no observable activity on Reddit at the time of retrieval. Of the remaining 147 accounts with activity, we excluded accounts with fewer than 10 combined active and passive interactions. 
This filtering step yielded a final set of 99 troll accounts, which we use as the positive class. 
The negative class of organic accounts was constructed from the dataset developed in \cite{yuan2025behavioral}, which was assembled through a multi-stage sampling procedure. 
That dataset targets activity across 15 subreddits spanning general news, political ideology, human rights, and sexual identity. For each subreddit, users were ranked by activity level, defined as the number of posts or comments within that subreddit. For each year from 2015 through 2021 (inclusive), we sampled up to 250 users per subreddit: the 50 most active users, plus additional users representing each remaining activity quartile so that lower-activity users were also included. When a subreddit had fewer than 250 active users in a given year, we included all of them. In total, this procedure produced 15{,}310 unique accounts.

\noindent From this pool of 15{,}310 users, we retained only those who were active between 2015 and 2018, in order to match the activity window of the trolls. This restriction resulted in 11{,}965 unique organic accounts. While the overall activity trajectory (\textit{i.e.}, the sequence and volume of observed actions, as well as the effective time window, which may be shortened for bots due to banning) differs substantially between trolls and organics over this interval, we explicitly account for trajectory length in several robustness experiments (see the \textit{Results} section).

\noindent Following \cite{yuan2025behavioral}, we fine-tune a pretrained DeBERTa-v3 model for agreement classification using the DEBAGREEMENT dataset \cite{pougue2021debagreement}. This dataset consists of comment-reply pairs from five subreddits: \textit{r/BlackLivesMatter}, \textit{r/Brexit}, \textit{r/climate}, \textit{r/democrats}, and \textit{r/Republican}. Each pair is labeled as ``agree,'' ``neutral,'' or ``disagree.'' This component is motivated by \cite{yuan2025behavioral}, who show that patterns of expressed agreement, disagreement, and neutrality capture core mechanisms of consensus-building in human discussion and reveal user groups with distinct behavioral homophily profiles.


\subsection*{Policy Inference}

\textit{Markov decision process.} User activity on social media platforms unfolds sequentially through interactions with both the platform and other users. Individuals display distinct behavioral patterns, ranging from passive consumption (\textit{e.g.}, lurkers) to highly active participation (\textit{e.g.}, influencers).  

\noindent Platforms provide a discrete set of interaction options, which in a sequential decision-making framework correspond to a set of actions $\mathcal{A}$. At any point in time $t$, we model the user's observable interaction context as a state $s_t \in \mathcal{S}$. User behavior is thus represented as the selection of an action $a_t \in \mathcal{A}$ conditional on the current state $s_t \in \mathcal{S}$. We assume that the next state depends only on the current state and action, consistent with the \emph{Markov property}. Accordingly, we represent user activity as a Markov Decision Process (MDP), defined by the tuple $(\mathcal{S}, \mathcal{A}, P, d_0, \gamma)$, where $\mathcal{S}$ is the set of states, $\mathcal{A}$ the set of actions, $P$ the state-transition kernel, $d_0$ the distribution over initial states, and $\gamma$ the discount factor. The state-transition kernel $P(s' \mid s,a)$ specifies the probability of moving to the next state $s' \in \mathcal{S}$ after taking action $a \in \mathcal{A}$ in state $s \in \mathcal{S}$, encoding user-platform dynamics. The trajectory $\tau$ records the complete sequence of user interactions $\tau = (s_0, a_0, \dots, s_{T-1}, a_{T-1}, s_T)$, where $T$ is the number of decision steps (\textit{i.e.}, the number of state-action pairs). From the trajectory of each user, we construct the \emph{empirical state-action visitation frequency} under policy $\pi$:  
\begin{equation}
\hat{\rho}_{\pi}(s,a) \coloneqq \frac{1}{T} \sum_{t=0}^{T-1} \mathds{1}{\{s_t = s,\, a_t = a \}}.
\end{equation}
The \emph{empirical user policy} $\hat{\pi}$ is then estimated from observed trajectories by normalizing visitation frequencies:  
\begin{equation}
\hat{\pi}(a \mid s) \coloneqq \frac{\hat{\rho}_{\pi}(s,a)}{\sum_{a' \in \mathcal{A}} \hat{\rho}_{\pi}(s,a')}.
\label{eq:empirical-policy}
\end{equation}
This empirical policy captures observable behavioral tendencies, which can, in principle, help distinguish malicious from benign users. Its effectiveness, however, depends on whether the observed trajectories---derived from publicly available platform data---are sufficiently informative. Recent restrictions on data access, driven by regulatory and operational changes, compound this challenge. For example, private messages or participation in non-public groups are typically excluded from datasets. Consequently, the definition of state and action spaces must be refined to reflect the information that is observable on a given platform. For X (formerly Twitter), prior work \cite{luceri2020detecting, geissler2023analyzing} proposes suitable encodings. For Reddit, we build on the encoding introduced in Yuan \textit{et al.}~\cite{yuan2025behavioral}, which we adopt and detail in the following section.
\smallskip

\noindent \textit{State and action space.} Following the reinforcement learning literature \cite{sutton1998reinforcement}, we model the user as an \emph{agent} operating within an \emph{environment}. The environment captures observable state changes on the Reddit platform, excluding actions that are unobservable (\textit{e.g.}, voting on content). It represents the entirety of the platform, excluding the focal user. Each user agent is therefore modeled independently and does not directly interact with other agents; interactions occur only indirectly, mediated through the environment.

\noindent A user's trajectory is constructed from the stream of events associated with that user across the platform, which we map into state-action feature pairs. We define the following state features:
\begin{itemize}[topsep=1pt]
\item \textit{Initial thread (IT).} First or only interaction, creating a new thread.
\item \textit{Initial root comment (IRC).} First or only interaction, posting a root comment.
\item \textit{Initial reply (IR).} First or only interaction, replying to a comment; split into 
\begin{itemize}[nosep, topsep=0pt, label=$\circ$]
    \item agreement ($IR_{+}$),
    \item neutrality ($IR_{\sim}$), and
    \item disagreement ($IR_{-}$).
\end{itemize}
\item \textit{Engaged root comment (ERC).} Already interacted, posting a root comment.
\item \textit{Engaged reply (ER).} Already interacted, replying to a comment; split into 
\begin{itemize}[nosep, topsep=0pt, label=$\circ$]
\item agreement ($ER_{+}$), 
\item neutrality ($ER_{\sim}$), 
\item disagreement ($ER_{-}$).
\end{itemize}
\item \textit{Get reply (GR).} Receiving a reply on any reply or comment; split into 
\begin{itemize}[nosep, topsep=0pt, label=$\circ$]
\item agreement ($GR_{+}$), 
\item neutrality ($GR_{\sim}$), 
\item disagreement ($GR_{-}$).
\end{itemize}
\end{itemize}
In summary, the environment consists of 12 states, with each agent beginning in one of five possible initial states
($\{IT, IRC, IR_{+}, IR_{\sim}, IR_{-}\}$).
At each timestep, after observing the current state, the agent selects one of six available actions, which determines the subsequent state transition.
\begin{itemize}[topsep=1pt]
\item \textit{Wait reply (WR).} User waits for a reply to one of their comments.
\item \textit{Create new thread (CT).} Start a new discussion thread in a subreddit.
\item \textit{Post root comment (RC).} Direct comment on thread's original post.
\item \textit{Post reply comment (PR).} Respond to another user's comment, creating a nested conversation. We further dissect this action between 
\begin{itemize}[nosep, topsep=0pt, label=$\circ$]
\item agreement ($PR_{+}$),
\item neutrality ($PR_{\sim}$),  
\item disagreement ($PR_{-}$).
\end{itemize}
\end{itemize}
Given the formal state and action space encoding, the selected user set, and the Pushshift Reddit API dataset, we extract per-user trajectories $\tau$, which form the basis for subsequent policy inference.
\smallskip

\noindent \textit{Policy learning.} Efficient policy learning depends on the dimensionality of the state and action spaces, as well as the length of individual user trajectories $\tau$. Information operations are often carried out by a set of coordinated agents. Depending on the platform, such agents attempt to evade detection by distributing activity across multiple accounts. As our goal is to identify malicious activity even from a limited number of observations, we require methods that remain robust in data-scarce regimes. The empirical policy (Eq.~\eqref{eq:empirical-policy}) provides a strong baseline when abundant data are available. However, it exhibits weak generalization to unseen states and suffers from high variance and sparsity.  

\noindent A simple alternative is \emph{imitation learning}, where policies are derived directly from state-action demonstrations. Behavioral cloning, the most common form, is computationally efficient but extrapolates poorly to novel states and ignores the long-term consequences of actions. Generative Adversarial Imitation Learning (GAIL) addresses these shortcomings by framing imitation as a two-player game: a discriminator distinguishes expert from learned trajectories, while the policy is trained to fool the discriminator. This adversarial setup encourages the policy to capture generalizable behavioral patterns rather than merely memorizing demonstrations.

\noindent \emph{Inverse Reinforcement Learning} (IRL) instead seeks to recover the latent reward function that explains observed behavior~\cite{ng2000,10.1145/279943.279964}. Because reward functions are not uniquely identifiable from behavior alone, regularization is required. Maximum entropy IRL imposes an entropy criterion to avoid degenerate or spurious reward assignments, ensuring that the inferred reward structure reflects plausible motivations behind observed actions. In summary, while the empirical policy amounts to memorization of observed behavior, GAIL and IRL aim to generalize to the underlying process dynamics, providing more reliable tools for distinguishing malicious from benign users.
\smallskip

\noindent \textit{Maximum Entropy Deep Inverse Reinforcement Learning.}
We formalize the behavior of an individual user in the framework of a Markov decision process (MDP). For a given user, the MDP is defined by the tuple $\mathcal{M} \coloneqq (\mathcal{S}, \mathcal{A}, P, d_0, \gamma)$, where $\mathcal{S}$ is a finite set of states, $\mathcal{A}$ a finite set of actions, $P(s' \mid s,a)$ the transition kernel specifying the probability of moving from state $s$ to $s'$ after taking action $a$, $d_0$ the distribution over initial states, and $\gamma \in (0,1)$ the discount factor. A stationary, stochastic policy $\pi(a \mid s)$ induces a distribution over the space of possible trajectories. The probability of a specific trajectory $\tau = (s_0,a_0,\dots,s_{T-1}, a_{T-1}, s_{T})$ with $T$ decision steps is given by the chain rule:
\begin{equation}
P_\pi(\tau) \coloneqq d_0(s_0)\,\prod_{t=0}^{T-1}\pi(a_t \mid s_t)\,P(s_{t+1} \mid s_t,a_t).
\end{equation}
To evaluate a policy, we define its \emph{discounted state visitation measure}
\begin{equation}
\mu_\pi(s) \coloneqq \sum_{t=0}^{T}\gamma^t\,\Pr(s_t = s \mid \pi, P),
\end{equation}
which describes the expected discounted frequency of visiting state $s$ under $\pi$. For a reward function $r:\mathcal{S}\to \mathbb{R}$, the expected discounted return is the inner product
\begin{equation}
\langle \mu_\pi, r\rangle \coloneqq \sum_{s \in \mathcal{S}} \mu_\pi(s)\,r(s)
= \mathbb{E}_{\tau \sim P_\pi}\!\Bigg[\sum_{t=0}^T \gamma^t r(s_t)\Bigg].
\end{equation}
Following the maximum entropy principle of Ziebart \textit{et al.}~\cite{ziebart2008maximum}, we regularize policies by their trajectory entropy $H(P_\pi) \coloneqq -\sum_{\tau} P_\pi(\tau)\,\log P_\pi(\tau)$,
which encourages stochasticity and prevents degenerate solutions. We parameterize the reward as a neural network $r_\theta$ with weights $\theta$ \cite{wulfmeier2015maximum}, and learn it by solving
\begin{equation}
\max_{r_\theta}\ \Big\{-\Psi(r_\theta)
+ \langle \mu_E, r_\theta \rangle
- \max_{\pi \in \Pi}\big[ \langle \mu_\pi, r_\theta \rangle + H(P_\pi) \big] \Big\},
\end{equation}
where $\Pi$ denotes the policy class, $\mu_E$ is the expert's discounted visitation measure estimated from demonstrations (summed over the observed finite trajectory), and $\Psi$ is a convex regularizer on the reward parameters.

\noindent In the model-based setting, the inner maximization and $\mu_\pi$ can be computed via dynamic programming under the known transition kernel $P$. The corresponding gradient update is
$\nabla_\theta\ \propto\ \sum_{s \in \mathcal{S}} \big(\mu_E(s) - \mu_\pi(s)\big)\,\nabla_\theta r_\theta(s)$,
which iteratively aligns the learner's visitation measure $\mu_\pi$ with that of the expert $\mu_E$. Given the learned reward $r_\theta$, we recover the policy using soft value iteration under the known dynamics, yielding a maximum-entropy policy that balances reward maximization and entropy via a temperature parameter. For further theoretical background and variants, see \cite{arora2021survey}.
\emph{Remark.} If the reward is defined on state-action pairs, $r:\mathcal{S}\times\mathcal{A}\to\mathbb{R}$, one replaces $\mu_\pi(s)$ by the discounted occupancy measure $\rho_\pi(s,a) \coloneqq \mu_\pi(s)\,\pi(a \mid s)$, and sums over $(s,a)$.
\smallskip

\noindent \textit{Generative Adversarial Imitation Learning.}  
Generative Adversarial Imitation Learning (GAIL) \cite{ho2016gail} bypasses explicit reward recovery and instead seeks to directly align the occupancy measure of the learned policy $\rho_\pi(s,a)$ with that of the expert $\rho_{\pi_E}(s,a)$. This is achieved by minimizing a divergence between the two, while retaining entropy regularization to promote stochasticity:  
\begin{equation}
\min_{\pi \in \Pi}\Big\{ D_{\mathrm{JS}}(\rho_\pi \,\Vert\, \rho_{\pi_E}) \;-\; \lambda\, H(P_\pi)\Big\},
\end{equation}
where $D_{\mathrm{JS}}$ denotes the Jensen-Shannon divergence, $H(P_\pi)$ is the trajectory entropy, and $\lambda>0$ controls the entropy preference.  

\noindent This problem can be expressed in adversarial form, introducing a discriminator $D:(\mathcal{S}\times\mathcal{A})\to(0,1)$:  
\begin{equation}
\min_{\pi}\;\sup_{D}\;
\Big\{
\mathbb{E}_{(s,a)\sim\rho_{\pi_E}}[\log D(s,a)]
+\mathbb{E}_{(s,a)\sim\rho_\pi}[\log(1-D(s,a))]
-\lambda\,H(P_\pi)
\Big\}.
\end{equation}
Here, the discriminator is trained to distinguish expert state-action pairs from those generated by $\pi$, while the policy is trained to deceive the discriminator. At equilibrium, the discriminator is no longer able to separate the two distributions, implying $\rho_\pi = \rho_{\pi_E}$. By parameterizing both the policy $\pi$ and discriminator $D$ with neural networks, GAIL supports flexible nonlinear policies that match long-horizon statistics of expert behavior. In practice, this approach achieves greater robustness than behavior cloning, especially when demonstrations are short or provide only sparse coverage of the state-action space.

\subsection*{Classification Framework}
\label{subsec:classification}

To classify users, we use Random Forest (RF) \cite{breiman2001rf} and eXtreme Gradient Boosting (XGBoost) \cite{chen2016xgboost}. Given a finite state and action space, each user's policy can be represented as a matrix of size $\lvert\mathcal{S}\rvert \times \lvert\mathcal{A}\rvert$, where $\lvert\mathcal{S}\rvert$ denotes the number of states and $\lvert\mathcal{A}\rvert$ the number of actions. Each entry corresponds to the probability of taking action $a \in \mathcal{A}$ given state $s \in \mathcal{S}$. In our setting, this yields a $12 \times 6$ matrix per user, which is vectorized and used as input to the classifier.

\noindent We adopt a matched case--control design rather than training directly on the full population due to the imbalance in positive and negative samples. In each run, every troll account is matched with one organic account based on trajectory length, yielding approximately 99 positive and 99 negative samples per run. This corresponds to controlled undersampling of the majority class.

\noindent Each experiment consists of 25 independent runs. Across runs, the set of troll accounts remains fixed, while the negative class is resampled, isolating the effect of variation in the control group. Within each run, we apply stratified $k$-fold cross-validation with $k=3$. Given the small sample size ($\sim$198 users per run), this ensures sufficient class support in each fold (approximately 66 positive examples for training and 33 for testing). Aggregating across runs and folds yields 75 out-of-fold evaluations per experiment configuration.

\subsection*{Evaluation Metrics}
\label{subsec:evaluation-metrics}
For our evaluation metrics in the binary classification setting (troll vs.~organic), we report the \emph{macro-}$F_{1}$ score (hereafter denoted simply as $F_{1}$). The macro-$F_{1}$ treats both classes $k \in \{1,\dots,K\}$ equally by averaging their per-class $F_{1}$ scores:
\begin{equation}
F_{1} \coloneqq \frac{1}{K}\sum_{k=1}^{K} F_{1,k}, 
\qquad 
F_{1,k} \coloneqq \frac{2\,\mathrm{Prec}_{k}\,\mathrm{Rec}_{k}}{\mathrm{Prec}_{k}+\mathrm{Rec}_{k}},
\end{equation}
where $F_{1,k}$ denotes the $F_{1}$ score of class $k$. Here, $\mathrm{Prec}_{k}$ and $\mathrm{Rec}_{k}$ denote the precision and recall of class $k$, respectively, defined from the confusion matrix entries---true positives ($\mathrm{TP}_{k}$), false positives ($\mathrm{FP}_{k}$), and false negatives ($\mathrm{FN}_{k}$)---as
\begin{equation}
\mathrm{Prec}_{k} \coloneqq \frac{\mathrm{TP}_{k}}{\mathrm{TP}_{k}+\mathrm{FP}_{k}}, 
\qquad 
\mathrm{Rec}_{k} \coloneqq \frac{\mathrm{TP}_{k}}{\mathrm{TP}_{k}+\mathrm{FN}_{k}}.
\end{equation}

\subsection*{Trajectory Perturbations: Noise and Hijacking}

\noindent\textit{Random Noise Perturbations.}
Let $\{(s_t, a_t)\}_{t=0}^{T-1}$ denote the state-action pairs extracted from a trajectory with $T$ decision steps, where $s_t \in \mathcal{S}$ and $a_t \in \mathcal{A}$, and let $p \in [0,1]$ represent the corruption rate.
We uniformly sample an index set
\begin{equation}
\mathcal{I} \subseteq \{0, 1, \dots, T-1\}, \quad \text{with} \quad \vert\mathcal{I}\vert = \lfloor p T \rfloor,
\end{equation}
where $\lfloor \cdot \rfloor$ denotes the floor function.
For each $t \in \mathcal{I}$, we replace $a_t$ by a value drawn uniformly from $\mathcal{A}$.
Because $s_{t+1}$ depends on $a_t$, we update $s_{t+1}$ by sampling uniformly from the set of legal next states $\mathcal{S}'(s_t,a_t)$.
If $t = 0$, we additionally resample $s_0$ uniformly from the valid set of initial states.
This procedure yields perturbed trajectories with randomized local deviations in behavior and state.
\smallskip

\noindent\textit{Account Hijacking Perturbations.}
We model an abrupt account hijacking attack by constructing synthetic trajectories in which behavior and content switch from an organic user profile to a troll profile at a specified time.

\noindent Let $T$ denote the trajectory length (in our experiments, $T = 100$), and let $\eta \in [0,1]$ be the fraction of the trajectory generated under an organic policy.
We define the switch point
\begin{equation}
\kappa \coloneqq \big\lfloor \eta T \big\rfloor.
\end{equation}
From the set of learned policies, we select one organic user $u^{\mathrm{org}}$ with policy $\pi^{\mathrm{org}}$ and one troll user $u^{\mathrm{troll}}$ with policy $\pi^{\mathrm{troll}}$.
Given the shared state space $\mathcal{S}$, action space $\mathcal{A}$, and environment dynamics $P(s' \mid s,a)$, we generate a hijacking trajectory $\tau^{\mathrm{hijack}} = \{(s_t,a_t)\}_{t=0}^{T-1}$ as
\begin{align}
a_t &\sim \pi_t(\cdot \mid s_t),
&
\pi_t &=
\begin{cases}
\pi^{\mathrm{org}}, & t < \kappa,\\[3pt]
\pi^{\mathrm{troll}}, & t \ge \kappa,
\end{cases}\\[6pt]
s_{t+1} &\sim P(\cdot \mid s_t, a_t),
& & t = 0,\dots,T-1.
\end{align}
This construction induces an abrupt change in behavioral policy at timestep $\kappa$, while remaining consistent with the learned platform dynamics.

\noindent To simulate the corresponding shift in content, we associate text embeddings with each generated action.
For each user $u$ and action $a \in \mathcal{A}$, let $\mathcal{C}_u^{a}$ denote the set of that user's historical content items produced under action $a$, and let $\mathcal{D}_u^{a}$ be the empirical distribution over their embeddings.
For each timestep $t$ in $\tau^{\mathrm{hijack}}$ with realized action $a_t$, we sample
\begin{equation}
c_t \sim \mathcal{D}_{u_t}^{a_t},
\qquad
u_t =
\begin{cases}
u^{\mathrm{org}}, & t < \kappa,\\[3pt]
u^{\mathrm{troll}}, & t \ge \kappa,
\end{cases}
\end{equation}
yielding a sequence of embeddings $\mathbf{c}_{0:T-1} = (c_0,\dots,c_{T-1})$ that mirrors the behavioral switch.
For content-based classification, we summarize this trajectory-level signal by the mean embedding
\begin{equation}
\bar{c} \coloneqq \frac{1}{T}\sum_{t=0}^{T-1} c_t,
\end{equation}
which serves as the trajectory-level representation in the hijacked-account scenario.

\noindent Organic user trajectories in this scenario are constructed by generating trajectories of length $T$ without introducing a shift.

\subsection*{Content Embeddings}
\label{subsec:content-embeddings}

For content-based detection, we represent each Reddit comment and submission as a fixed-dimensional embedding vector. We use ModernBERT-large \cite{warner2024smarter} with pretrained weights and the reference implementation as a frozen encoder. ModernBERT is an encoder-only Transformer architecture that incorporates recent advances in pretraining and model design, trained on approximately 2 trillion tokens with a native context length of 8{,}192 tokens. It achieves state-of-the-art performance on a broad suite of classification and retrieval benchmarks while remaining computationally efficient, making it well suited for large-scale encoding of heterogeneous Reddit text.

\noindent We apply the same preprocessing to troll and organic users. Comments are encoded directly from their body text. Submissions are encoded from the concatenation of the title and the self-text field when available; for link-only submissions, we use the title alone. Entries with missing, empty, or deleted text are discarded prior to encoding.

\noindent All texts are tokenized with the ModernBERT-large tokenizer. Sequences longer than 512 tokens are truncated to the first 512 tokens, and shorter sequences are padded within each batch. We then extract the contextualized representation of the \texttt{[CLS]} token from the final hidden layer as the content embedding. This procedure yields a 1{,}024-dimensional vector for each remaining item, which serves as the base representation for downstream content-based classifiers.

\section*{Data availability}
Instructions for obtaining the source data are available at:
\url{https://github.com/behavioral-ds/behavioral-dynamics-io}.

\section*{Code availability}
Code for reproducing the analyses is available at:
\url{https://github.com/behavioral-ds/behavioral-dynamics-io}.

\section*{Acknowledgements}
This research was supported by the Australian Government Research Training Program (RTP) Scholarship; the Commonwealth of Australia (represented by the Defence Science and Technology Group) through a Defence Science Partnerships Agreement; the Australian Department of Home Affairs; the Advanced Strategic Capabilities Accelerator (ASCA); the Defence Innovation Network; and the Australian Academy of Science.
This research was undertaken with the assistance of resources and services from the National Computational Infrastructure (NCI), which is supported by the Australian Government.

\section*{Author Contributions}
P.J.S. and L.Y. contributed equally to this work. P.J.S. and L.Y.: Conceptualization; Methodology; Investigation; Formal analysis; Validation; Data curation; Writing - Original Draft; Writing - Review \& Editing. M.-A.R.: Conceptualization; Supervision; Funding acquisition; Writing - Review \& Editing.

\section*{Competing Interests}
The authors declare no competing interests.

\startsupplement

\begin{center}

\vspace*{0.8cm}

{\sffamily\Large\bfseries Supplementary Information}

\vspace{0.9cm}

{\LARGE\bfseries Beyond Content: Behavioral Policies Reveal Actors in Information Operations}

\vspace{1.15cm}

{\large
Philipp J. Schneider$^{1,*,\dagger}$,
Lanqin Yuan$^{2,*}$,
Marian-Andrei Rizoiu$^{2}$
}

\vspace{0.5cm}

{\footnotesize
$^{1}$College of Management of Technology,
\'{E}cole Polytechnique F\'{e}d\'{e}rale de Lausanne,
CH-1015 Lausanne, Switzerland
}

\vspace{0.1cm}

{\footnotesize
$^{2}$Faculty of Engineering and Information Technology,
University of Technology Sydney,
Ultimo, NSW 2007, Australia
}

\vspace{0.5cm}

{\footnotesize
$^{*}$These authors contributed equally to this work.\\
$^{\dagger}$Correspondence: \href{mailto:philipp.schneider@epfl.ch}{philipp.schneider@epfl.ch}
}

\end{center}

\vspace{0.5em}


\section*{Overview}

This Supplementary Information provides additional material supporting the dataset construction, behavioral representations, and empirical analyses reported in the main manuscript.

\medskip

\noindent Section~S1 provides supplementary information on the Reddit dataset and behavioral policy representations. Section~S2 reports additional experimental results, including baseline comparisons, robustness analyses, model-specific results, and hyperparameter selection procedures.

\vspace{1em}

\tableofcontents

\clearpage


\section{Dataset}
\label{app:dataset}
\subsection{Descriptive Statistics}
\begin{figure}[!htbp]
    \centering
    \begin{subfigure}[b]{0.48\textwidth}
        \centering
        \includegraphics[width=\textwidth]{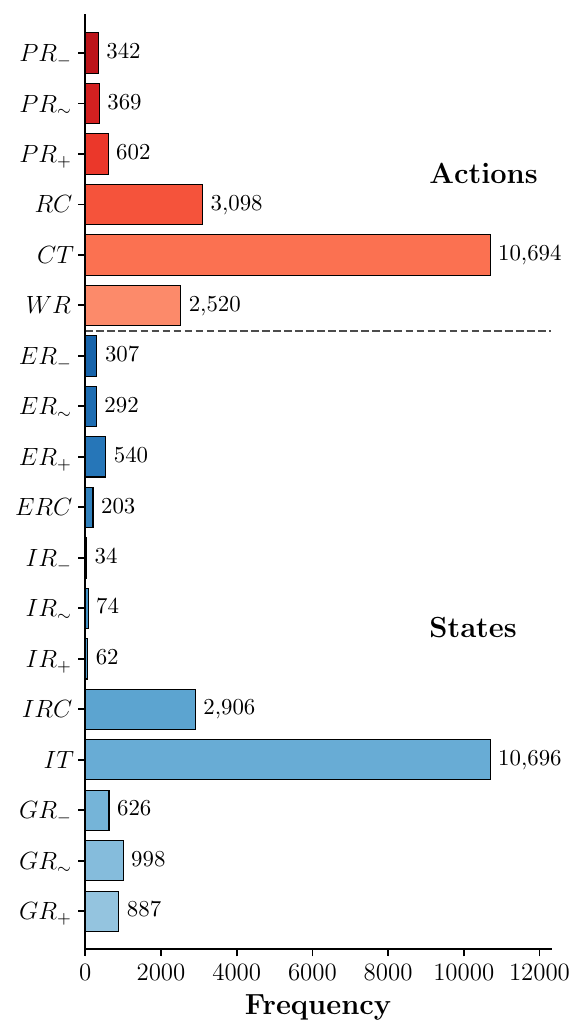}
        \caption{Trolls.}
        \label{fig:sa-distributions-trolls}
    \end{subfigure}
    \hfill
    \begin{subfigure}[b]{0.48\textwidth}
        \centering
        \includegraphics[width=\textwidth]{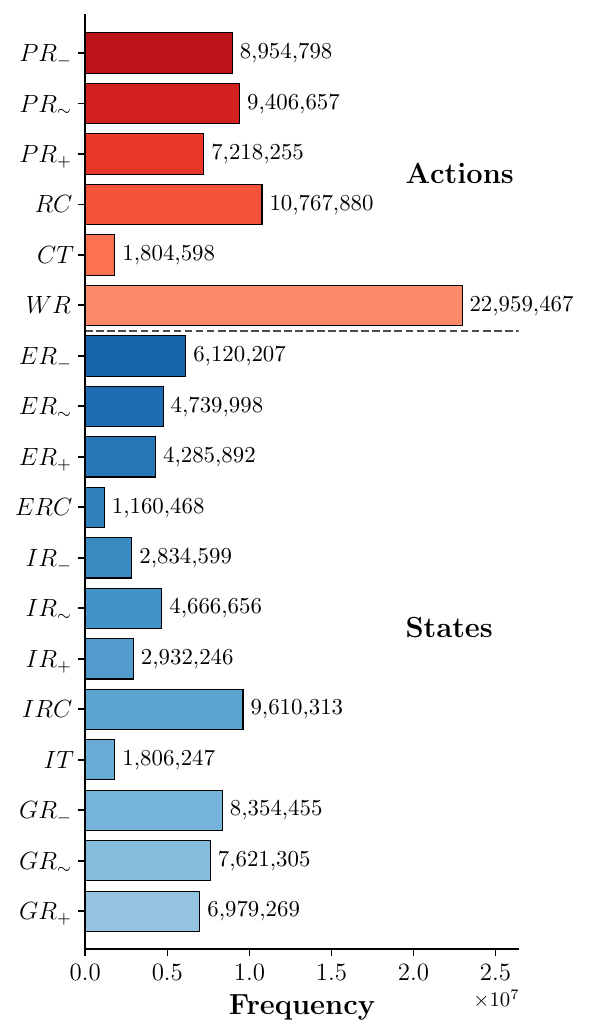}
        \caption{Organics.}
        \label{fig:sa-distributions-organics}
    \end{subfigure}
    \caption{Action and state frequencies across user groups.}
    \label{fig:action-state-distribution}
\end{figure}

This section expands upon the dataset description in \textit{Methods}, providing further insight into the action and state distributions for both troll and organic accounts. Figure~\ref{fig:action-state-distribution} visualizes these distributions for all 99 troll accounts (see also Fig.~\ref{fig:sa-distributions-trolls}) and a matched sample of 11,965 organic accounts (Fig.~\ref{fig:sa-distributions-organics}). As expected, the overall volume of actions and visited states reflects the class imbalance between trolls and organics. However, qualitative differences in behavioral focus emerge. In particular, reply actions---namely $PR_{-}$ (negative), $PR_{\sim}$ (neutral), and $PR_{+}$ (positive)---are more prevalent among organic users, reflecting a tendency to engage in ongoing conversations. In contrast, troll accounts more frequently initiate discussions via thread creation ($CT$), suggesting a strategic intent to steer or seed discourse rather than participate organically.

\noindent Table~\ref{tab:troll_subreddits} lists the most frequently targeted subreddits by troll accounts. These communities span a diverse range of themes, including news, politics, cryptocurrency, memes, and NSFW content, highlighting the broad operational scope of coordinated influence efforts.

\begin{table}[htbp]
\caption{Top 20 Most Frequent Subreddits for Troll Accounts}
\label{tab:troll_subreddits}
\begin{tabular}{lrr}
\toprule
Subreddit & Count & Percentage ($\%$) \\
\midrule
\textit{r/uncen} & $1776$ & $11.3$ \\
\textit{r/Bad\_Cop\_No\_Donut} & 860 & $5.47$ \\
\textit{r/AskReddit} & $814$ & $5.18$ \\
\textit{r/CryptoCurrency} & $687$ & $4.37$ \\
\textit{r/funny} & $570$ & $3.63$ \\
\textit{r/POLITIC} & $306$ & $1.95$ \\
\textit{r/racism} & $301$ & $1.92$ \\
\textit{r/news} & $284$ & $1.81$ \\
\textit{r/worldnews} & $269$ & $1.71$ \\
\textit{r/aww} & $246$ & $1.57$ \\
\textit{r/politics} & $232$ & $1.48$ \\
\textit{r/PoliticalHumor} & $222$ & $1.41$ \\
\textit{r/copwatch} & $214$ & $1.36$ \\
\textit{r/gifs} & $210$ & $1.34$ \\
\textit{r/blackpower} & $205$ & $1.3$ \\
\textit{r/Bitcoin} & $197$ & $1.25$ \\
\textit{r/uspolitics} & $176$ & $1.12$ \\
\textit{r/interestingasfuck} & $176$ & $1.12$ \\
\textit{r/The\_Donald} & $164$ & $1.04$ \\
\textit{r/police} & $163$ & $1.04$ \\
\bottomrule
\end{tabular}
\end{table}

\subsection{Low-Dimensional Embeddings of Behavioral Policies}
As shown in Results, policy-based encodings of user activity outperform content-based approaches in identifying troll accounts. To further examine how individual user policies are organized within the high-dimensional policy space, we apply two popular manifold learning techniques---\emph{t-distributed Stochastic Neighbor Embedding} (t-SNE) \cite{vandermaaten08a} and \emph{Uniform Manifold Approximation and Projection} (UMAP) \cite{mcinnes2018umap}---to visualize the pairwise distances between users based on their empirical policies.

\begin{figure}[!htbp] 
    \centering
    \includegraphics[width=0.6\textwidth]{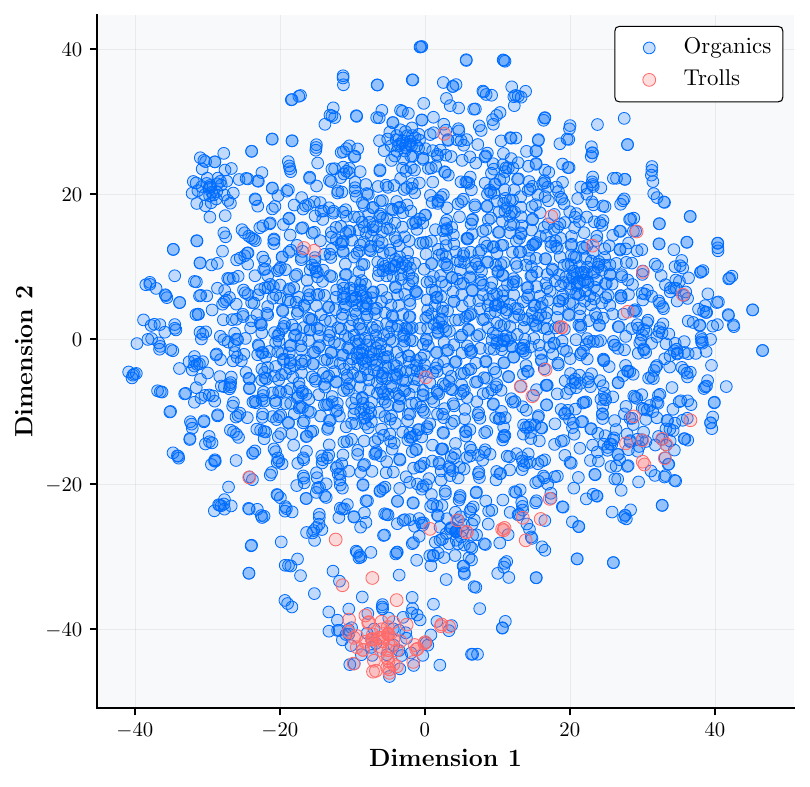} 
    \caption{t-SNE projection using Euclidean distance between user policies.}
    \label{fig:tsne-euclid}  
\end{figure}

\begin{figure}[!htbp] 
    \centering
    \includegraphics[width=0.6\textwidth]{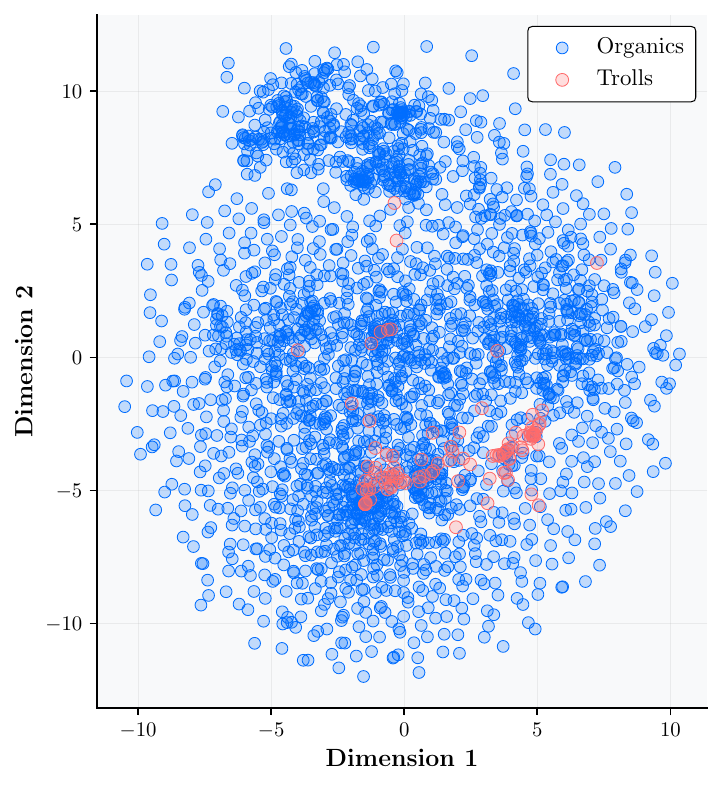}
    \caption{t-SNE projection using SWKL between user policies.}
    \label{fig:tsne-swkl}

\end{figure}

\begin{figure}[!htbp] 
    \centering
    \includegraphics[width=0.6\textwidth]{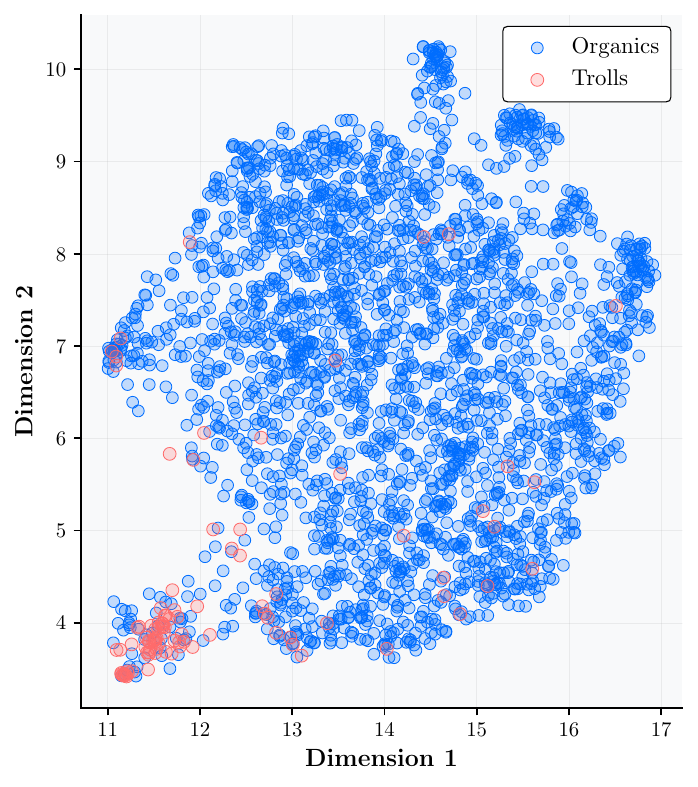} 
    \caption{UMAP projection using Euclidean distance between user policies.}
    \label{fig:umap-euclid}  
\end{figure}

\begin{figure}[!htbp] 
    \centering
    \includegraphics[width=0.9\textwidth]{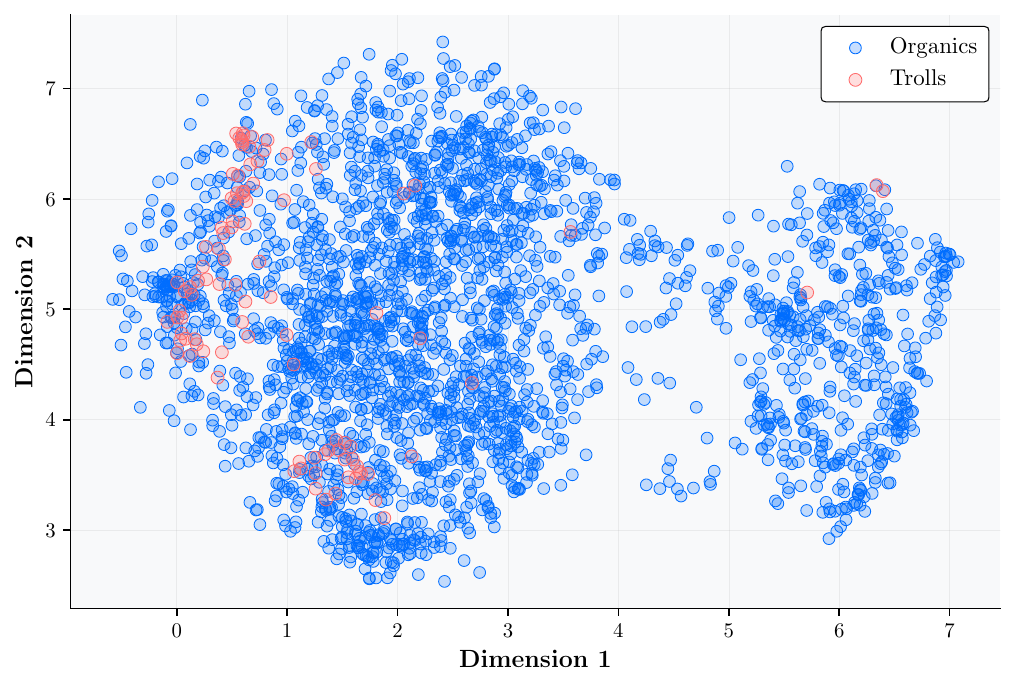}
    \caption{UMAP projection using SWKL between user policies.}
    \label{fig:umap-swkl}
\end{figure}

\noindent We evaluate two distance metrics: standard Euclidean distance and the symmetric weighted KL-divergence (SWKL) recently introduced in \cite{yuan2025behavioral}. These metrics are applied to the action distributions inferred from each user's behavioral trajectory. The resulting two-dimensional projections are presented in Figures~\ref{fig:tsne-euclid}, \ref{fig:tsne-swkl}, \ref{fig:umap-euclid}, and \ref{fig:umap-swkl}.

\noindent Across all configurations, we find no clear separation between troll and organic users in the embedded space. This suggests that while the learned policies are effective for classification in high dimensions, they do not form linearly separable clusters in two dimensions---underscoring the behavioral overlap and complexity observed in real-world user populations.

\pagebreak

\section{Supplementary Experimental Results}
\label{app:supp-results}

\subsection{Comparison with Prior Work}

Table~\ref{tab:sota-comparison} summarizes closely related account-level detection studies. It highlights key differences in evaluation settings, dataset construction, and performance metrics across studies, which limit direct numerical comparison (see the \textit{Results} section for further discussion).

\begin{sidewaystable}[p]
    \centering
    \caption{Summary of closely related account-level detection studies. Performance values are reported as stated in the original publications; substantial differences in evaluation settings limit direct comparisons across studies.}
    \footnotesize
    \setlength{\tabcolsep}{2.5pt}
    \begin{tabular}{p{2.5cm} p{1.5cm} p{3.5cm} p{4.0cm} p{1.2cm} p{1.4cm} p{4.0cm}}
        \toprule
        Study & Platform & Target class & Negative class & Metric & Value & Comparability\\
        \midrule
        Luceri \textit{et al.}~\cite{luceri2020detecting}        & Twitter  & IRA-linked trolls & Election-discussion accounts (Twitter users collected via 2016 US election keywords; topically matched) & AUC & $89.1\%$ & Different platform; negative class not matched on activity volume \\
        Kong \textit{et al.}~\cite{kong2023interval}          & Twitter  & State-sponsored accounts (three countries: Russia, Iran, Saudi Arabia; Twitter Moderation Research Consortium) & Users in same retweet cascades, not labeled as state-sponsored & Macro-$F_1$ & $98.7\%$ & Different platform; negatives defined by cascade co-participation, not matched on activity or trajectory length \\
        Alizadeh \textit{et al.}~\cite{alizadeh2020content}      & Twitter \& Reddit & State-sponsored accounts (Chinese, Russian, Venezuelan on Twitter; Russian/IRA on Reddit) & Random U.S.\ accounts \& politically engaged users (Twitter); posts from political subreddits (Reddit) & Macro-$F_1$ & $82\%$ (Reddit campaign) & Content-based; Reddit campaign directly comparable; within-month evaluation \\
        Saeed \textit{et al.}~\cite{saeed2022}         & Reddit   & Russian-sponsored trolls (2017 transparency report) & Random Reddit accounts from same subreddits as known trolls, excluding suspended accounts & $F_1$ & $97.8\%$ & Same source dataset as this work; different filtering; negative class not trajectory-matched; detection relies on cross-account behavioral similarity \\
        Cresci \textit{et al.}~\cite{cresci2017social} (DDNA) & Twitter  & Social spambots (political retweeters + Amazon product spammers) & Researcher-verified human accounts (crowdsensing, 3{,}474 accounts) & MCC & $95.5\%$ & Different platform; bot class differs from state-sponsored trolls \\
        Nwala \textit{et al.}~\cite{nwala2023language} (BLOC)  & Twitter  & Social bots (14 annotated datasets, Bot Repository) & Human-labeled accounts pooled across Bot Repository datasets (Twitter-verified, celebrity, and manually annotated users) & $F_1$ & $89.2\%$ & Different platform; Reddit-adapted in the present study \\
        \midrule
        \textbf{This work}                    & Reddit   & IRA-linked trolls (2017 transparency report) & Matched organic accounts (1:1 on trajectory length)            & Macro-$F_1$ & $94.9\%$ (GAIL) & Matched case-control; policy-based, behavior-only (no platform-specific metadata)\\
        \bottomrule
    \end{tabular}
    \label{tab:sota-comparison}
\end{sidewaystable}

\subsection{Activity-Based Baselines}

To benchmark our approach, we compare against two activity-based baselines that represent user behavior as symbolic sequences: digital DNA (DDNA) \cite{cresci2017social} and the Behavioral Languages for Online Characterization (BLOC) framework \cite{nwala2023language}. Both methods encode user activity as sequences over a finite alphabet and can therefore be adapted to Reddit's interaction structure.
\smallskip

\noindent \textbf{Digital DNA (DDNA).}
DDNA represents user activity as sequences of symbols over a predefined action alphabet and measures similarity between users based on longest common subsequence (LCS) patterns. We construct a minimal alphabet consisting of directly observable user actions: creation of a root thread ($CT$), commenting on a root thread ($RC$), and replying to an existing comment ($PR$).

\noindent For each user, we construct a symbolic sequence ordered by time. Pairwise similarity between sequences is computed using LCS-based measures, which capture shared subsequence structure while allowing for gaps. These similarity features are subsequently used for classification under the same experimental protocol as the policy-based representations.
\smallskip

\noindent \textbf{Behavioral Languages for Online Characterization (BLOC).}
BLOC encodes user activity as symbolic sequences constructed from an action alphabet and, in its original formulation, a content alphabet. As Reddit does not provide features such as hashtags, mentions, or explicit user-user links, we restrict the representation to the action alphabet and omit the content alphabet.

\noindent We construct a Reddit-compatible action alphabet using the same observable actions as in DDNA ($CT$, $RC$, $PR$). To capture temporal dynamics, we augment the sequences with coarse inter-event time tokens inserted between consecutive actions. Specifically, for a time difference $t$ between two actions, we insert:
\begin{itemize}[itemsep=0pt, parsep=0pt, topsep=0pt, partopsep=0pt]
    \item $m$: $1$ minute $< t < 1$ hour,
    \item $h$: $1$ hour $< t < 1$ day,
    \item $d$: $1$ day $< t < 1$ week,
    \item $w$: $t > 1$ week.
\end{itemize}

\noindent The resulting sequences are encoded using character-level $n$-grams (uni-, bi-, and tri-grams) with TF-IDF weighting, yielding fixed-length feature vectors for each user. These vectors are used as input to standard classifiers (Random Forest and XGBoost).
\smallskip

\noindent \textbf{Evaluation Protocol.} Both baselines are evaluated under the same protocol as the policy-based models to ensure comparability. In particular, user trajectories are truncated to the first $N$ actions for early-detection analysis, with $N$ ranging from 3 to 50. For each representation, features are constructed from the truncated sequences and used for classification under the matched case--control design and cross-validation procedure described in the \textit{Methods} (Classification Framework) section.

\begin{figure}[!htbp]
\includegraphics[width=1.0\textwidth]{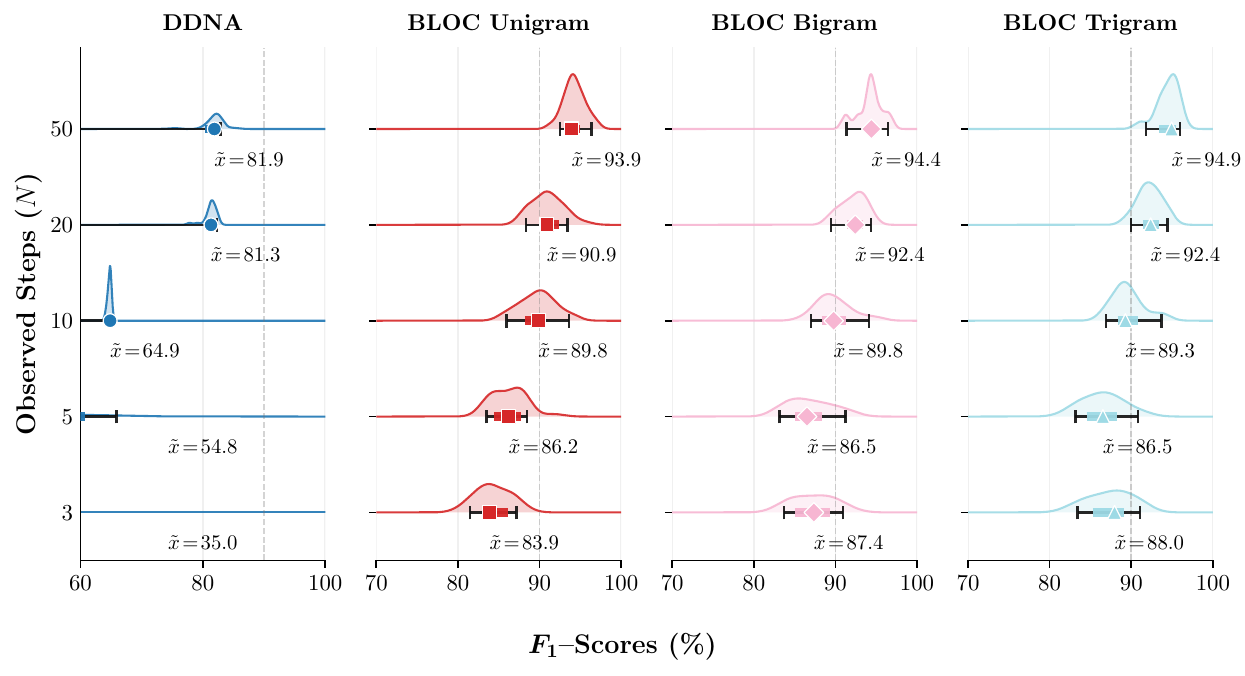} 
\caption{\textbf{Classification performance by trajectory length (DDNA and BLOC baselines)}. 
Median macro-$F_1$ (with 5th-95th percentiles) when classifiers are trained on the first $n$ actions, with $n$ ranging from 3 to 50 observations.}
\label{fig:traj-length-ddna-bloc}
\end{figure}

\noindent We present additional results for the DDNA and BLOC baselines in Fig.~\ref{fig:traj-length-ddna-bloc}. DDNA is evaluated using the three-action encoding ($CT$, $RC$, $PR$), while for BLOC we report uni-gram, bi-gram, and tri-gram representations.
\smallskip

\noindent \textbf{Results.} Across experiments, policy-based classifiers consistently outperform the DDNA baseline. This gap is most pronounced at short trajectory lengths, reflecting the reliance of DDNA on longest common subsequence patterns, which require sufficiently long sequences to extract stable similarity signals.

\noindent A similar pattern is observed for BLOC. Policy-based approaches (GAIL, IRL, and empirical policies) outperform BLOC in the low-data regime, while BLOC performance improves with increasing trajectory length and begins to approach that of the policy-based models for longer sequences (\textit{e.g.}, $N \approx 20$). This suggests that aggregated timing and action-frequency patterns become informative only once sufficient observations are available.

\noindent This distinction is particularly relevant in practice, where user actions arrive sequentially and decisions must be made from partial histories. The strong performance of policy-based representations at short trajectories therefore translates directly into earlier detection, whereas DDNA and BLOC require longer observation windows to accumulate comparable statistical signal.

\subsection{Detailed Classification Metrics}

Tables~\ref{tab:method_policy_metrics_n10_rf} and~\ref{tab:method_activity_metrics_n10_rf} report detailed per-class classification metrics (precision, recall, and $F_1$-score) at $N=10$ observations for the policy- and content-based methods and the activity-based baselines (DDNA, BLOC), respectively. These metrics provide a more fine-grained view of class-specific performance and help clarify differences in precision and recall across organic users and trolls.

\begin{table*}[!t]
\centering
\small
\setlength{\tabcolsep}{4pt}
\caption{Per-class classification metrics at $N=10$ (Random Forest) for policy- and content-based methods. Metrics are reported in \%.}
\label{tab:method_policy_metrics_n10_rf}
\begin{tabular}{lrrrrrrrrrrr}
\toprule
Method & $TN$ & $FP$ & $FN$ & $TP$ & $P_0$ & $R_0$ & $F_{1,0}$ & $P_1$ & $R_1$ & $F_{1,1}$ & Macro-$F_1$ (runs) \\
\midrule
Empirical & $2184$ & $291$ & $231$ & $2244$ & $90.4$ & $88.2$ & $89.3$ & $88.5$ & $90.7$ & $89.6$ & $89.5\,\pm\,1.8$ \\
GAIL & $2264$ & $211$ & $276$ & $2199$ & $89.1$ & $91.5$ & $90.3$ & $91.2$ & $88.8$ & $90.0$ & $90.2\,\pm\,1.6$ \\
MaxEnt Deep IRL & $2223$ & $252$ & $262$ & $2213$ & $89.5$ & $89.8$ & $89.6$ & $89.8$ & $89.4$ & $89.6$ & $89.6\,\pm\,1.5$ \\
Embedding & $2048$ & $427$ & $435$ & $1940$ & $82.5$ & $82.7$ & $82.6$ & $82.0$ & $81.7$ & $81.8$ & $82.2\,\pm\,2.5$ \\
\bottomrule
\end{tabular}
\parbox{\textwidth}{\footnotesize\raggedright
Class labels are encoded as $0=\text{organic}$ and $1=\text{troll}$. 
$P_k$, $R_k$, and $F_{1,k}$ denote the precision, recall, and class-specific $F_1$-score for class $k \in \{0,1\}$.}
\end{table*}

\begin{table*}[t]
\centering
\small
\setlength{\tabcolsep}{4pt}
\caption{Per-class classification metrics at $N=10$ (Random Forest) for activity-based baselines. Metrics are reported in \%.}
\label{tab:method_activity_metrics_n10_rf}
\begin{tabular}{lrrrrrrrrrrr}
\toprule
Method & $TN$ & $FP$ & $FN$ & $TP$ & $P_0$ & $R_0$ & $F_{1,0}$ & $P_1$ & $R_1$ & $F_{1,1}$ & Macro-$F_1$ (runs) \\
\midrule
DDNA & $2311$ & $145$ & $1548$ & $902$ & $59.9$ & $94.1$ & $73.2$ & $86.2$ & $36.8$ & $51.6$ & $62.4\,\pm\,4.8$ \\
BLOC Unigram & $2192$ & $264$ & $238$ & $2212$ & $90.2$ & $89.3$ & $89.7$ & $89.3$ & $90.3$ & $89.8$ & $89.8\,\pm\,2.2$ \\
BLOC Bigram & $2207$ & $249$ & $249$ & $2201$ & $89.9$ & $89.9$ & $89.9$ & $89.8$ & $89.8$ & $89.8$ & $89.8\,\pm\,2.1$ \\
BLOC Trigram & $2213$ & $243$ & $268$ & $2182$ & $89.2$ & $90.1$ & $89.6$ & $90.0$ & $89.1$ & $89.5$ & $89.6\,\pm\,1.9$ \\
\bottomrule
\end{tabular}
\parbox{\textwidth}{\footnotesize\raggedright
Class labels are encoded as $0=\text{organic}$ and $1=\text{troll}$. 
$P_k$, $R_k$, and $F_{1,k}$ denote the precision, recall, and class-specific $F_1$-score for class $k \in \{0,1\}$.}
\end{table*}

\subsection{Classification Results for eXtreme Gradient Boosting}
\begin{figure}[!htbp] 
    \centering
    \includegraphics[width=0.9\textwidth]{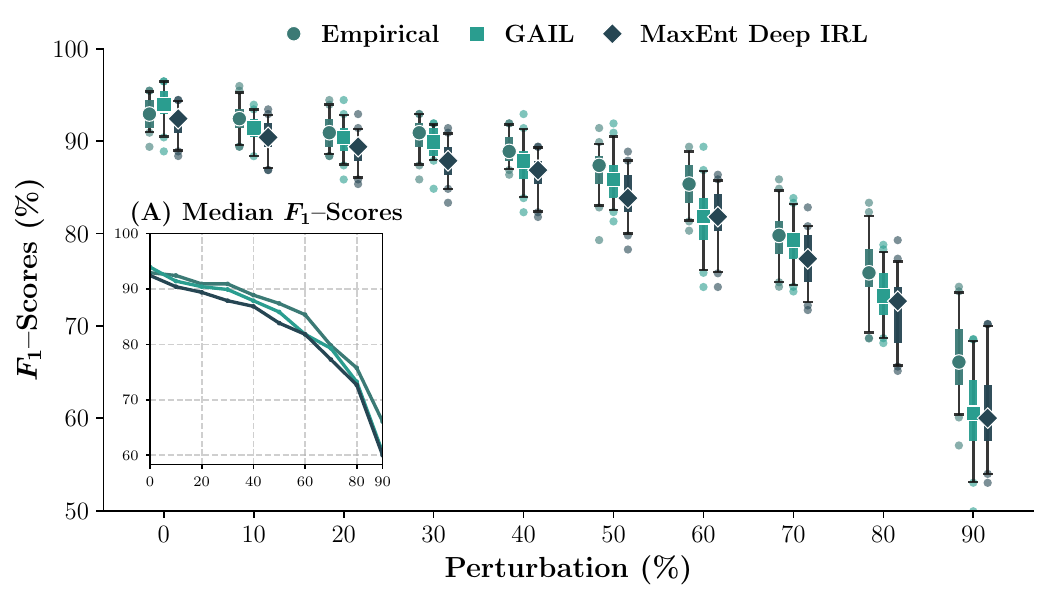} 
    \caption{\textbf{Performance on perturbed trajectories (XGBoost).} Median macro-$F_1$ for Empirical, IRL and GAIL as a function of the perturbation level $p$ (fraction of state-action pairs replaced); inset (A) shows the median $F_{1}$-scores across methods.}
    \label{fig:perturb-xgb}  
\end{figure}

\begin{figure}[!htbp] 
    \centering
    \includegraphics[width=1.0\textwidth]{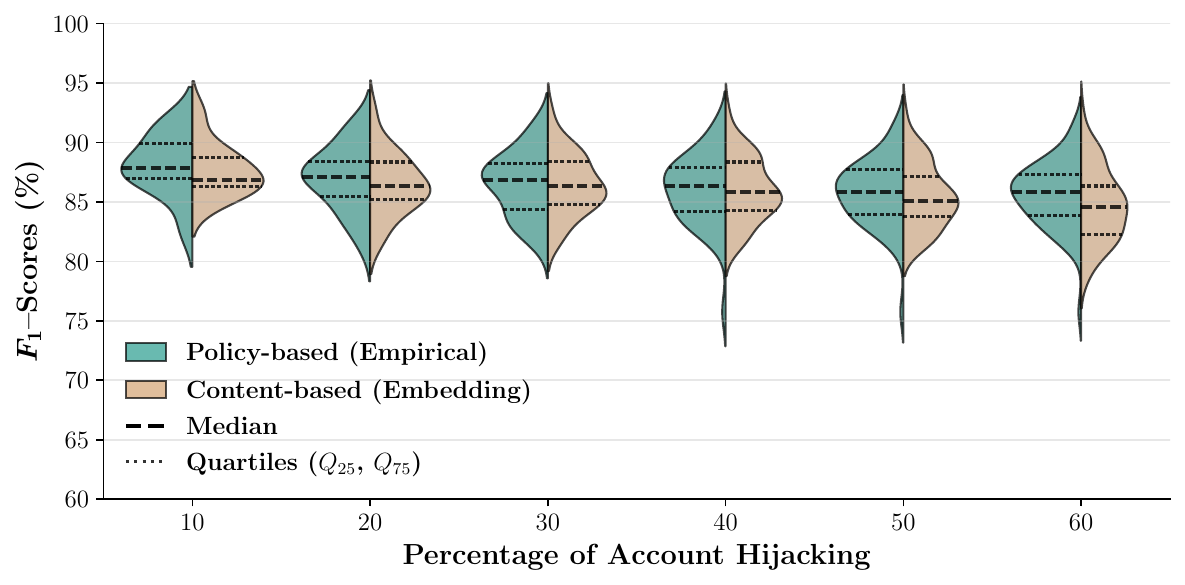}
    \caption{\textbf{Account hijacking (XGBoost).} Policy-based vs content-based classification performance by severity of hijacking.}
    \label{fig:simulate-mixed-policy-xg}
\end{figure}

In the \textit{Results} section, we identified the Random Forest (RF) classifier as achieving superior classification performance---measured by macro $F_1$-score---compared to eXtreme Gradient Boosting (XGBoost). For completeness, we report the results obtained using XGBoost in Figures~\ref{fig:perturb-xgb} and \ref{fig:simulate-mixed-policy-xg}.

\noindent Figure~\ref{fig:perturb-xgb} mirrors the trends observed in Figure~2 (C): classification performance degrades as a larger portion of trajectories are perturbed. Consistent with prior results, GAIL outperforms empirical policies in the unperturbed setting but exhibits sharper performance decline as perturbations increase.

\noindent Similarly, Figure~\ref{fig:simulate-mixed-policy-xg} confirms that policy-based models consistently outperform content-based approaches across varying levels of account hijacking, replicating the trend seen with the RF classifier in Figure~3 (C).

\subsection{Clusters of Troll Behavior}
\begin{figure}[!htbp]
    \centering
    \begin{subfigure}[b]{0.45\textwidth}
        \centering
        \includegraphics[width=\textwidth]{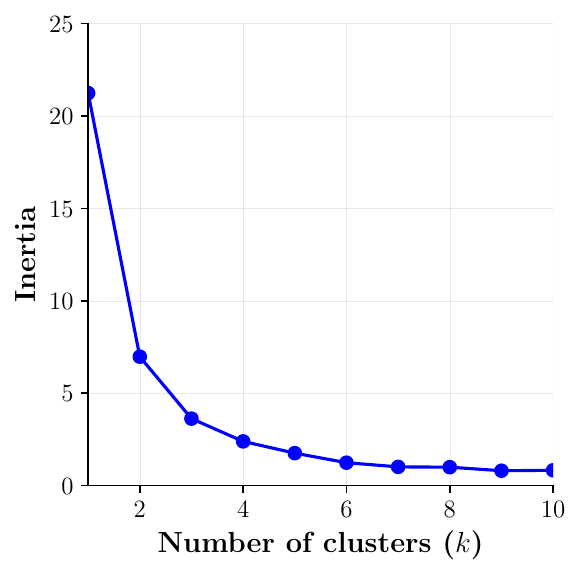}
        \caption{Elbow method.}
        \label{fig:cluster-elbow}
    \end{subfigure}
    \hfill
    \begin{subfigure}[b]{0.45\textwidth}
        \centering
        \includegraphics[width=\textwidth]{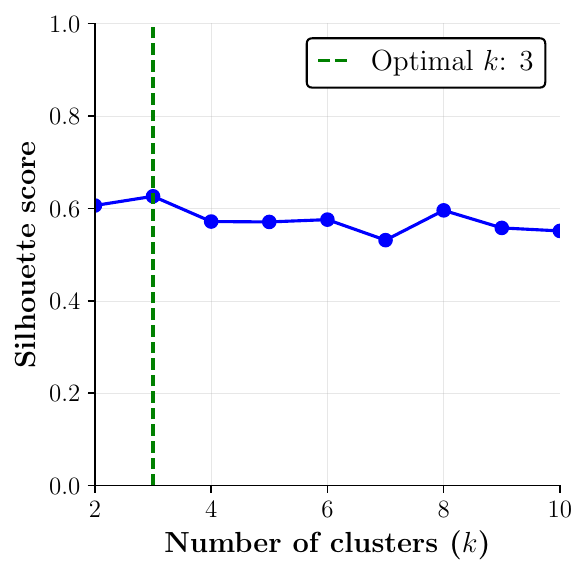}
        \caption{Silhouette score.}
        \label{fig:cluster-silhouette-score}
    \end{subfigure}
    \caption{Cluster selection for $k$-means.}
    \label{fig:cluster-selection}
\end{figure}

To determine the appropriate number of clusters for the $k$‑means algorithm, we applied two complementary selection criteria: the silhouette score and the elbow method. Figure~\ref{fig:cluster-selection} reports these metrics for varying values of $k$ using the troll sample. Based on the relative changes observed in both criteria, we selected $k = 3$ as the final clustering solution.

\noindent We also qualitatively examined the results for $k = 2$ and $k = 4$. The $k = 2$ solution yields an overly coarse partition, distinguishing only between a group of trolls that almost exclusively create threads and the remaining accounts. In contrast, $k = 4$ produces clusters that closely resemble those at $k = 3$, with minor refinements that capture more niche behavioral variants. We therefore adopt $k = 3$ as it captures meaningful variation without introducing unnecessary complexity.

\newpage

\subsection{Action Distributions of Troll and Organic Users}

To further examine group-level behavioral differences, we marginalize the GAIL-derived policies over the state space to obtain empirical action distributions for both troll and organic users. This approach summarizes each account's behavior by the relative frequency of actions---such as thread creation, root-level comments, and replies---aggregated across all observed states.

\begin{figure}[!htbp]
    \centering
    \includegraphics[width=0.6\linewidth]{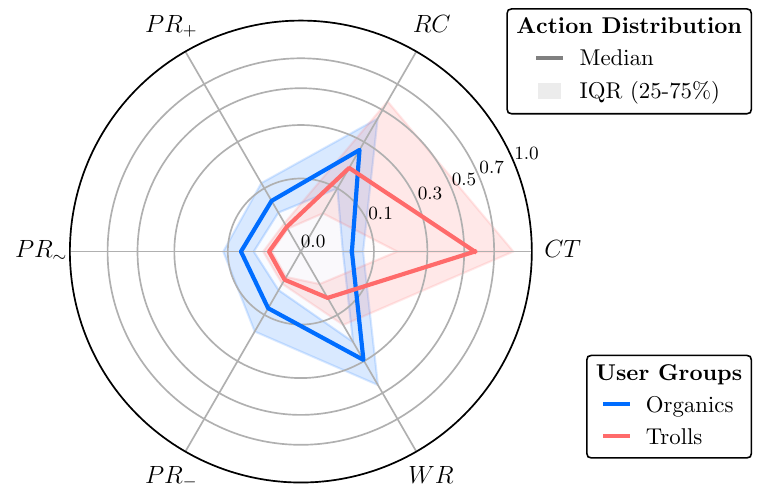}
    \caption{Empirical action distributions of troll and organic users, aggregated over states.}
    \label{fig:radar-organics-trolls-gail}
\end{figure}

\begin{figure}[!htbp]
    \centering
    \includegraphics[width=0.6\linewidth]{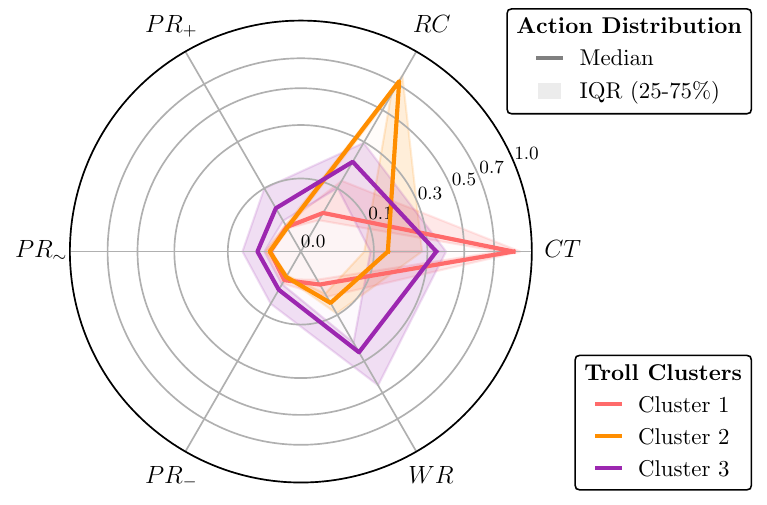}
    \caption{Empirical action distributions of the $k=3$ troll clusters.}
    \label{fig:radar-troll-clusters-gail}
\end{figure}

\begin{figure}[!htbp]
    \centering
    \includegraphics[width=0.6\linewidth]{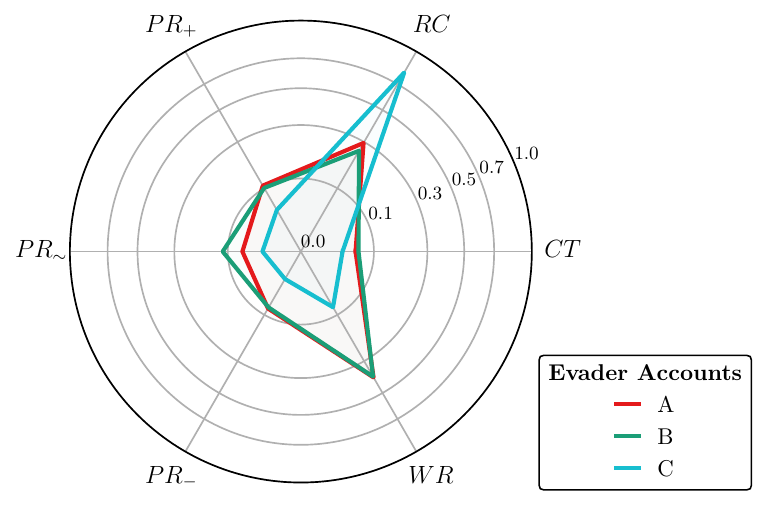}
    \caption{Empirical action distributions of the three users who consistently evaded detection.}
    \label{fig:radar-evaders-gail}
\end{figure}

\noindent Figure~\ref{fig:radar-organics-trolls-gail} reveals marked differences in how trolls and organic users engage with Reddit discussions. We further visualize the behavioral archetypes uncovered by our clustering analysis (Fig.~3 (A)) via corresponding action distributions in Figure~\ref{fig:radar-troll-clusters-gail}. Finally, Figure~\ref{fig:radar-evaders-gail} shows the action profiles of three accounts that consistently evaded detection, offering insight into deceptive behaviors that closely mimic organic usage patterns.


\newpage

\subsection{Hyperparameter Search for Maximum-Entropy Deep IRL}

Classification performance is highly sensitive to the hyperparameter choices used during policy inference. In this section, we detail the hyperparameter search for maximum-entropy deep inverse reinforcement learning (IRL), as outlined in Table~\ref{tab:deep-irl-hyperparams}. The corresponding results are reported in Tables~\ref{tab:medirl-lr-005}--\ref{tab:medirl-lr-05}, covering different learning rates $\alpha$. To guide optimal model selection, we assess each configuration using macro $F_1$-scores from both Random Forest (RF) and eXtreme Gradient Boosting (XGBoost) classifiers.

\begin{table}[htbp]
\caption{Hyperparameters for Maximum-Entropy Deep IRL.}
\small
\begin{tabular}{lc}
\toprule
\textbf{Hyperparameter} & \textbf{Value}\\
\midrule
Learning rate ($\alpha$) & \{0.005, 0.01, 0.05\}\\
Epochs & \{500, 1{,}000, 1{,}500\}\\
Discount factor ($\gamma$) & \{0.9, 0.95\}\\
Convergence threshold ($\epsilon$) & 0.01\\
Weight initialization ($w$) & Normal\\
Optimizer & Adam \\
Neural network structure & (12, 3, 3)\\
\bottomrule
\end{tabular}
\label{tab:deep-irl-hyperparams}
\end{table}

\begin{table}
\caption{Maximum-Entropy Deep IRL: Classification performance at learning rate $\alpha = 0.005$.}
\label{tab:medirl-lr-005}
\scriptsize
\begin{tabular}{rrrrcccc}
\toprule
 & & & & \multicolumn{2}{c}{$F_1$-Score} & \multicolumn{2}{c}{Recall}\\ \cmidrule(lr){5-6} \cmidrule(lr){7-8}
$\gamma$ & Epochs & $\lambda_1$ & $\lambda_2$ & RF & XGBoost & RF & XGBoost \\
\midrule
\multirow{9}{*}{$0.90$}& \multirow{9}{*}{$500$} & \multirow{3}{*}{$0.0$} & $0.0$ & $0.893 \pm 0.008$ & $0.898 \pm 0.006$ & $0.859 \pm 0.017$ & $0.893 \pm 0.015$ \\
& & & $0.5$ & $0.885 \pm 0.014$ & $0.884 \pm 0.017$ & $0.855 \pm 0.012$ & $0.867 \pm 0.017$ \\
& & & $1.0$ & $0.877 \pm 0.014$ & $0.859 \pm 0.015$ & $0.834 \pm 0.022$ & $0.838 \pm 0.011$ \\
& & \multirow{3}{*}{$0.5$} & $0.0$ & $0.868 \pm 0.014$ & $0.868 \pm 0.020$ & $0.824 \pm 0.019$ & $0.844 \pm 0.012$ \\
& & & $0.5$ & $0.861 \pm 0.026$ & $0.858 \pm 0.024$ & $0.808 \pm 0.034$ & $0.838 \pm 0.018$ \\
& & & $1.0$ & $0.853 \pm 0.025$ & $0.846 \pm 0.019$ & $0.804 \pm 0.038$ & $0.820 \pm 0.024$ \\
& & \multirow{3}{*}{$1.0$} & $0.0$ & $0.864 \pm 0.021$ & $0.864 \pm 0.019$ & $0.818 \pm 0.025$ & $0.844 \pm 0.008$ \\
& & & $0.5$ & $0.864 \pm 0.022$ & $0.852 \pm 0.025$ & $0.816 \pm 0.031$ & $0.826 \pm 0.024$ \\
& & & $1.0$ & $0.854 \pm 0.022$ & $0.846 \pm 0.017$ & $0.806 \pm 0.024$ & $0.824 \pm 0.014$ \\
\midrule
\multirow{9}{*}{$0.90$} & \multirow{9}{*}{$1000$} & \multirow{3}{*}{$0.0$} & $0.0$ & $0.916 \pm 0.005$ & $0.914 \pm 0.012$ & $0.897 \pm 0.010$ & $0.907 \pm 0.017$ \\
& & & $0.5$ & $0.896 \pm 0.010$ & $0.874 \pm 0.017$ & $0.871 \pm 0.017$ & $0.865 \pm 0.005$ \\
& & & $1.0$ & $0.892 \pm 0.012$ & $0.880 \pm 0.009$ & $0.859 \pm 0.011$ & $0.863 \pm 0.005$ \\
& & \multirow{3}{*}{$0.5$} & $0.0$ & $0.853 \pm 0.021$ & $0.863 \pm 0.022$ & $0.798 \pm 0.032$ & $0.844 \pm 0.027$ \\
& & & $0.5$ & $0.858 \pm 0.020$ & $0.860 \pm 0.019$ & $0.812 \pm 0.019$ & $0.840 \pm 0.017$ \\
& & & $1.0$ & $0.856 \pm 0.019$ & $0.862 \pm 0.027$ & $0.812 \pm 0.024$ & $0.853 \pm 0.019$ \\
& & \multirow{3}{*}{$1.0$} & $0.0$ & $0.851 \pm 0.018$ & $0.858 \pm 0.021$ & $0.802 \pm 0.028$ & $0.836 \pm 0.013$ \\
& & & $0.5$ & $0.854 \pm 0.023$ & $0.863 \pm 0.029$ & $0.808 \pm 0.026$ & $0.842 \pm 0.015$ \\
& & & $1.0$ & $0.853 \pm 0.030$ & $0.853 \pm 0.021$ & $0.806 \pm 0.030$ & $0.836 \pm 0.021$ \\
\midrule
\multirow{9}{*}{$0.90$} & \multirow{9}{*}{$1500$} & \multirow{3}{*}{$0.0$} & $0.0$ & $0.932 \pm 0.009$ & $0.930 \pm 0.007$ & $0.921 \pm 0.013$ & $0.929 \pm 0.011$ \\
& & & $0.5$ & $0.906 \pm 0.010$ & $0.888 \pm 0.009$ & $0.873 \pm 0.015$ & $0.877 \pm 0.015$ \\
& & & $1.0$ & $0.883 \pm 0.010$ & $0.884 \pm 0.019$ & $0.840 \pm 0.025$ & $0.867 \pm 0.027$ \\
& & \multirow{3}{*}{$0.5$} & $0.0$ & $0.881 \pm 0.021$ & $0.871 \pm 0.016$ & $0.844 \pm 0.024$ & $0.861 \pm 0.021$ \\
& & & $0.5$ & $0.857 \pm 0.014$ & $0.854 \pm 0.019$ & $0.808 \pm 0.018$ & $0.830 \pm 0.016$ \\
& & & $1.0$ & $0.860 \pm 0.009$ & $0.845 \pm 0.017$ & $0.806 \pm 0.017$ & $0.820 \pm 0.021$ \\
& & \multirow{3}{*}{$1.0$} & $0.0$ & $0.851 \pm 0.008$ & $0.845 \pm 0.017$ & $0.802 \pm 0.014$ & $0.812 \pm 0.026$ \\
& & & $0.5$ & $0.854 \pm 0.007$ & $0.846 \pm 0.027$ & $0.802 \pm 0.019$ & $0.820 \pm 0.031$ \\
& & & $1.0$ & $0.858 \pm 0.010$ & $0.841 \pm 0.016$ & $0.804 \pm 0.020$ & $0.804 \pm 0.024$ \\
\midrule
\multirow{9}{*}{$0.95$} & \multirow{9}{*}{$500$} & \multirow{3}{*}{$0.0$} & $0.0$ & $0.903 \pm 0.013$ & $0.897 \pm 0.020$ & $0.873 \pm 0.014$ & $0.889 \pm 0.014$ \\
& & & $0.5$ & $0.888 \pm 0.017$ & $0.884 \pm 0.014$ & $0.848 \pm 0.018$ & $0.867 \pm 0.025$ \\
& & & $1.0$ & $0.873 \pm 0.011$ & $0.857 \pm 0.010$ & $0.826 \pm 0.017$ & $0.826 \pm 0.020$ \\
& & \multirow{3}{*}{$0.5$} & $0.0$ & $0.863 \pm 0.009$ & $0.855 \pm 0.018$ & $0.814 \pm 0.018$ & $0.826 \pm 0.023$ \\
& & & $0.5$ & $0.847 \pm 0.022$ & $0.843 \pm 0.021$ & $0.790 \pm 0.023$ & $0.812 \pm 0.012$ \\
& & & $1.0$ & $0.841 \pm 0.020$ & $0.834 \pm 0.021$ & $0.778 \pm 0.028$ & $0.800 \pm 0.017$ \\
& & \multirow{3}{*}{$1.0$} & $0.0$ & $0.851 \pm 0.010$ & $0.851 \pm 0.010$ & $0.794 \pm 0.019$ & $0.816 \pm 0.022$ \\
& & & $0.5$ & $0.837 \pm 0.020$ & $0.838 \pm 0.011$ & $0.782 \pm 0.024$ & $0.806 \pm 0.020$ \\
& & & $1.0$ & $0.833 \pm 0.016$ & $0.823 \pm 0.020$ & $0.776 \pm 0.012$ & $0.790 \pm 0.022$ \\
\midrule
\multirow{9}{*}{$0.95$} & \multirow{9}{*}{$1000$} & \multirow{3}{*}{$0.0$} & $0.0$ & $0.916 \pm 0.009$ & $0.915 \pm 0.014$ & $0.891 \pm 0.012$ & $0.905 \pm 0.022$ \\
& & & $0.5$ & $0.893 \pm 0.018$ & $0.883 \pm 0.010$ & $0.863 \pm 0.014$ & $0.873 \pm 0.019$ \\
& & & $1.0$ & $0.879 \pm 0.015$ & $0.877 \pm 0.015$ & $0.836 \pm 0.012$ & $0.859 \pm 0.011$ \\
& & \multirow{3}{*}{$0.5$} & $0.0$ & $0.856 \pm 0.019$ & $0.855 \pm 0.010$ & $0.800 \pm 0.028$ & $0.836 \pm 0.017$ \\
& & & $0.5$ & $0.857 \pm 0.015$ & $0.847 \pm 0.015$ & $0.800 \pm 0.013$ & $0.824 \pm 0.029$ \\
& & & $1.0$ & $0.858 \pm 0.016$ & $0.854 \pm 0.009$ & $0.798 \pm 0.013$ & $0.830 \pm 0.027$ \\
& & \multirow{3}{*}{$1.0$} & $0.0$ & $0.852 \pm 0.013$ & $0.854 \pm 0.014$ & $0.788 \pm 0.019$ & $0.830 \pm 0.020$ \\
& & & $0.5$ & $0.855 \pm 0.012$ & $0.857 \pm 0.015$ & $0.794 \pm 0.012$ & $0.828 \pm 0.029$ \\
& & & $1.0$ & $0.851 \pm 0.007$ & $0.852 \pm 0.010$ & $0.790 \pm 0.016$ & $0.824 \pm 0.022$ \\
\midrule
\multirow{9}{*}{$0.95$} & \multirow{9}{*}{$1500$} & \multirow{3}{*}{$0.0$} & $0.0$ & $0.930 \pm 0.015$ & $0.929 \pm 0.015$ & $0.925 \pm 0.018$ & $0.923 \pm 0.019$ \\
& & & $0.5$ & $0.910 \pm 0.005$ & $0.899 \pm 0.011$ & $0.885 \pm 0.014$ & $0.891 \pm 0.016$ \\
& & & $1.0$ & $0.884 \pm 0.011$ & $0.884 \pm 0.014$ & $0.842 \pm 0.018$ & $0.863 \pm 0.008$ \\
& & \multirow{3}{*}{$0.5$} & $0.0$ & $0.874 \pm 0.018$ & $0.873 \pm 0.015$ & $0.840 \pm 0.022$ & $0.863 \pm 0.008$ \\
& & & $0.5$ & $0.849 \pm 0.007$ & $0.852 \pm 0.019$ & $0.794 \pm 0.014$ & $0.828 \pm 0.024$ \\
& & & $1.0$ & $0.853 \pm 0.016$ & $0.856 \pm 0.017$ & $0.802 \pm 0.010$ & $0.836 \pm 0.008$ \\
& & \multirow{3}{*}{$1.0$} & $0.0$ & $0.844 \pm 0.007$ & $0.849 \pm 0.018$ & $0.790 \pm 0.020$ & $0.822 \pm 0.012$ \\
& & & $0.5$ & $0.839 \pm 0.009$ & $0.846 \pm 0.015$ & $0.786 \pm 0.004$ & $0.826 \pm 0.013$ \\ 
& & & $1.0$ & $0.841 \pm 0.009$ & $0.846 \pm 0.020$ & $0.782 \pm 0.014$ & $0.808 \pm 0.032$ \\
\bottomrule
\end{tabular}
\end{table}

\clearpage

\begin{table}[htbp]
\caption{Maximum-Entropy Deep IRL: Classification performance at learning rate $\alpha = 0.01$.}
\label{tab:medirl-lr-01}
\scriptsize
\begin{tabular}{rrrrcccc}
\toprule
 & & & & \multicolumn{2}{c}{$F_1$-Score} & \multicolumn{2}{c}{Recall}\\ \cmidrule(lr){5-6} \cmidrule(lr){7-8}
$\gamma$ & Epochs & $\lambda_1$ & $\lambda_2$ & RF & XGBoost & RF & XGBoost \\
\midrule
\multirow{9}{*}{$0.90$} & \multirow{9}{*}{$500$} & \multirow{3}{*}{$0.0$} & $0.0$ & $0.933 \pm 0.004$ & $0.928 \pm 0.021$ & $0.923 \pm 0.005$ & $0.929 \pm 0.021$ \\
& & & $0.5$ & $0.908 \pm 0.010$ & $0.912 \pm 0.009$ & $0.875 \pm 0.014$ & $0.891 \pm 0.010$ \\
& & & $1.0$ & $0.898 \pm 0.018$ & $0.894 \pm 0.020$ & $0.865 \pm 0.020$ & $0.863 \pm 0.024$ \\
& & \multirow{3}{*}{$0.5$} & $0.0$ & $0.890 \pm 0.008$ & $0.885 \pm 0.005$ & $0.855 \pm 0.010$ & $0.879 \pm 0.018$ \\
& & & $0.5$ & $0.873 \pm 0.009$ & $0.855 \pm 0.024$ & $0.832 \pm 0.024$ & $0.836 \pm 0.013$ \\
& & & $1.0$ & $0.860 \pm 0.014$ & $0.854 \pm 0.023$ & $0.812 \pm 0.016$ & $0.826 \pm 0.023$ \\
& & \multirow{3}{*}{$1.0$} & $0.0$ & $0.876 \pm 0.012$ & $0.866 \pm 0.018$ & $0.832 \pm 0.027$ & $0.842 \pm 0.023$ \\
& & & $0.5$ & $0.861 \pm 0.005$ & $0.841 \pm 0.016$ & $0.812 \pm 0.014$ & $0.820 \pm 0.008$ \\
& & & $1.0$ & $0.859 \pm 0.007$ & $0.838 \pm 0.021$ & $0.810 \pm 0.010$ & $0.818 \pm 0.017$ \\
\midrule
\multirow{9}{*}{$0.90$} & \multirow{9}{*}{$1000$} & \multirow{3}{*}{$0.0$} & $0.0$ & $0.935 \pm 0.010$ & $0.928 \pm 0.021$ & $0.925 \pm 0.008$ & $0.927 \pm 0.022$ \\
& & & $0.5$ & $0.914 \pm 0.008$ & $0.897 \pm 0.016$ & $0.897 \pm 0.008$ & $0.885 \pm 0.016$ \\
& & & $1.0$ & $0.908 \pm 0.014$ & $0.910 \pm 0.017$ & $0.877 \pm 0.025$ & $0.895 \pm 0.008$ \\
& & \multirow{3}{*}{$0.5$} & $0.0$ & $0.916 \pm 0.011$ & $0.907 \pm 0.016$ & $0.905 \pm 0.012$ & $0.893 \pm 0.023$ \\
& & & $0.5$ & $0.906 \pm 0.010$ & $0.909 \pm 0.012$ & $0.889 \pm 0.014$ & $0.901 \pm 0.020$ \\
& & & $1.0$ & $0.897 \pm 0.015$ & $0.900 \pm 0.017$ & $0.875 \pm 0.018$ & $0.887 \pm 0.020$ \\
& & \multirow{3}{*}{$1.0$} & $0.0$ & $0.905 \pm 0.009$ & $0.907 \pm 0.016$ & $0.889 \pm 0.013$ & $0.909 \pm 0.014$ \\
& & & $0.5$ & $0.891 \pm 0.020$ & $0.898 \pm 0.018$ & $0.869 \pm 0.023$ & $0.899 \pm 0.009$ \\
& & & $1.0$ & $0.887 \pm 0.015$ & $0.896 \pm 0.016$ & $0.851 \pm 0.023$ & $0.899 \pm 0.021$ \\
\midrule
\multirow{9}{*}{$0.90$} & \multirow{9}{*}{$1500$} & \multirow{3}{*}{$0.0$} & $0.0$ & $0.932 \pm 0.009$ & $0.933 \pm 0.020$ & $0.923 \pm 0.008$ & $0.929 \pm 0.018$ \\
& & & $0.5$ & $0.918 \pm 0.015$ & $0.921 \pm 0.012$ & $0.909 \pm 0.014$ & $0.905 \pm 0.019$ \\
& & & $1.0$ & $0.914 \pm 0.010$ & $0.906 \pm 0.019$ & $0.897 \pm 0.010$ & $0.887 \pm 0.015$ \\
& & \multirow{3}{*}{$0.5$} & $0.0$ & $0.913 \pm 0.012$ & $0.918 \pm 0.013$ & $0.887 \pm 0.023$ & $0.901 \pm 0.010$ \\
& & & $0.5$ & $0.919 \pm 0.008$ & $0.916 \pm 0.015$ & $0.897 \pm 0.012$ & $0.905 \pm 0.014$ \\
& & & $1.0$ & $0.915 \pm 0.012$ & $0.910 \pm 0.021$ & $0.901 \pm 0.013$ & $0.899 \pm 0.019$ \\
& & \multirow{3}{*}{$1.0$} & $0.0$ & $0.920 \pm 0.008$ & $0.910 \pm 0.012$ & $0.889 \pm 0.017$ & $0.893 \pm 0.010$ \\
& & & $0.5$ & $0.920 \pm 0.014$ & $0.912 \pm 0.013$ & $0.903 \pm 0.019$ & $0.907 \pm 0.026$ \\
& & & $1.0$ & $0.907 \pm 0.014$ & $0.902 \pm 0.022$ & $0.893 \pm 0.015$ & $0.899 \pm 0.031$ \\
\midrule
\multirow{9}{*}{$0.95$} & \multirow{9}{*}{$500$} & \multirow{3}{*}{$0.0$} & $0.0$ & $0.930 \pm 0.013$ & $0.933 \pm 0.009$ & $0.921 \pm 0.015$ & $0.925 \pm 0.016$ \\
& & & $0.5$ & $0.910 \pm 0.008$ & $0.899 \pm 0.014$ & $0.881 \pm 0.004$ & $0.885 \pm 0.016$ \\
& & & $1.0$ & $0.892 \pm 0.008$ & $0.881 \pm 0.011$ & $0.855 \pm 0.016$ & $0.859 \pm 0.013$ \\
& & \multirow{3}{*}{$0.5$} & $0.0$ & $0.885 \pm 0.017$ & $0.897 \pm 0.010$ & $0.857 \pm 0.028$ & $0.891 \pm 0.017$ \\
& & & $0.5$ & $0.866 \pm 0.013$ & $0.850 \pm 0.018$ & $0.830 \pm 0.021$ & $0.842 \pm 0.029$ \\
& & & $1.0$ & $0.853 \pm 0.007$ & $0.847 \pm 0.014$ & $0.800 \pm 0.020$ & $0.836 \pm 0.017$ \\
& & \multirow{3}{*}{$1.0$} & $0.0$ & $0.865 \pm 0.015$ & $0.850 \pm 0.020$ & $0.832 \pm 0.025$ & $0.826 \pm 0.017$ \\
& & & $0.5$ & $0.842 \pm 0.014$ & $0.837 \pm 0.013$ & $0.794 \pm 0.026$ & $0.822 \pm 0.016$ \\
& & & $1.0$ & $0.843 \pm 0.008$ & $0.827 \pm 0.011$ & $0.792 \pm 0.014$ & $0.800 \pm 0.021$ \\
\midrule
\multirow{9}{*}{$0.95$} & \multirow{9}{*}{$1000$} & \multirow{3}{*}{$0.0$} & $0.0$ & $0.941 \pm 0.014$ & $0.934 \pm 0.016$ & $0.933 \pm 0.016$ & $0.935 \pm 0.014$ \\
& & & $0.5$ & $0.912 \pm 0.013$ & $0.910 \pm 0.010$ & $0.899 \pm 0.013$ & $0.895 \pm 0.010$ \\
& & & $1.0$ & $0.910 \pm 0.014$ & $0.900 \pm 0.010$ & $0.893 \pm 0.010$ & $0.891 \pm 0.004$ \\
& & \multirow{3}{*}{$0.5$} & $0.0$ & $0.916 \pm 0.014$ & $0.906 \pm 0.019$ & $0.899 \pm 0.018$ & $0.895 \pm 0.028$ \\
& & & $0.5$ & $0.907 \pm 0.014$ & $0.899 \pm 0.023$ & $0.901 \pm 0.008$ & $0.893 \pm 0.020$ \\
& & & $1.0$ & $0.896 \pm 0.020$ & $0.887 \pm 0.016$ & $0.867 \pm 0.017$ & $0.879 \pm 0.017$ \\
& & \multirow{3}{*}{$1.0$} & $0.0$ & $0.908 \pm 0.008$ & $0.896 \pm 0.014$ & $0.889 \pm 0.013$ & $0.889 \pm 0.019$ \\
& & & $0.5$ & $0.897 \pm 0.020$ & $0.890 \pm 0.014$ & $0.883 \pm 0.024$ & $0.887 \pm 0.016$ \\
& & & $1.0$ & $0.888 \pm 0.018$ & $0.891 \pm 0.015$ & $0.855 \pm 0.029$ & $0.887 \pm 0.008$ \\
\midrule
\multirow{9}{*}{$0.95$} & \multirow{9}{*}{$1500$} & \multirow{3}{*}{$0.0$} & $0.0$ & $0.936 \pm 0.013$ & $0.939 \pm 0.020$ & $0.925 \pm 0.018$ & $0.933 \pm 0.022$\\
& & & $0.5$ & $0.925 \pm 0.014$ & $0.935 \pm 0.009$ & $0.907 \pm 0.015$ & $0.933 \pm 0.012$ \\
& & & $1.0$ & $0.919 \pm 0.008$ & $0.911 \pm 0.024$ & $0.907 \pm 0.015$ & $0.899 \pm 0.027$ \\
& & \multirow{3}{*}{$0.5$} & $0.0$ & $0.915 \pm 0.006$ & $0.914 \pm 0.016$ & $0.887 \pm 0.010$ & $0.893 \pm 0.024$ \\
& & & $0.5$ & $0.916 \pm 0.008$ & $0.910 \pm 0.016$ & $0.897 \pm 0.010$ & $0.905 \pm 0.015$ \\
& & & $1.0$ & $0.914 \pm 0.014$ & $0.907 \pm 0.005$ & $0.899 \pm 0.020$ & $0.911 \pm 0.012$ \\
& & \multirow{3}{*}{$1.0$} & $0.0$ & $0.927 \pm 0.010$ & $0.921 \pm 0.012$ & $0.899 \pm 0.017$ & $0.913 \pm 0.019$ \\
& & & $0.5$ & $0.911 \pm 0.011$ & $0.903 \pm 0.012$ & $0.897 \pm 0.015$ & $0.895 \pm 0.024$ \\
& & & $1.0$ & $0.904 \pm 0.014$ & $0.892 \pm 0.013$ & $0.883 \pm 0.016$ & $0.891 \pm 0.015$ \\
\bottomrule
\end{tabular}
\end{table}

\clearpage

\begin{table}[htbp]
\caption{Maximum-Entropy Deep IRL: Classification performance at learning rate $\alpha = 0.05$.}
\label{tab:medirl-lr-05}
\scriptsize
\begin{tabular}{rrrrcccc}
\toprule
 & & & & \multicolumn{2}{c}{$F_1$-Score} & \multicolumn{2}{c}{Recall}\\ \cmidrule(lr){5-6} \cmidrule(lr){7-8}
$\gamma$ & Epochs & $\lambda_1$ & $\lambda_2$ & RF & XGBoost & RF & XGBoost \\
\midrule
\multirow{9}{*}{$0.90$} & \multirow{9}{*}{$500$} & \multirow{3}{*}{$0.0$} & $0.0$ & $0.936 \pm 0.011$ & $0.934 \pm 0.019$ & $0.931 \pm 0.013$ & $0.931 \pm 0.016$ \\
& & & $0.5$ & $0.931 \pm 0.016$ & $0.932 \pm 0.021$ & $0.925 \pm 0.015$ & $0.925 \pm 0.023$ \\
& & & $1.0$ & $0.926 \pm 0.018$ & $0.924 \pm 0.020$ & $0.917 \pm 0.017$ & $0.913 \pm 0.019$ \\
& & \multirow{3}{*}{$0.5$} & $0.0$ & $0.930 \pm 0.017$ & $0.931 \pm 0.016$ & $0.927 \pm 0.008$ & $0.929 \pm 0.013$ \\
& & & $0.5$ & $0.919 \pm 0.017$ & $0.912 \pm 0.009$ & $0.915 \pm 0.024$ & $0.905 \pm 0.019$ \\
& & & $1.0$ & $0.936 \pm 0.015$ & $0.923 \pm 0.016$ & $0.921 \pm 0.016$ & $0.917 \pm 0.015$ \\
& & \multirow{3}{*}{$1.0$} & $0.0$ & $0.907 \pm 0.016$ & $0.897 \pm 0.022$ & $0.899 \pm 0.011$ & $0.901 \pm 0.013$ \\
& & & $0.5$ & $0.924 \pm 0.016$ & $0.917 \pm 0.026$ & $0.911 \pm 0.022$ & $0.907 \pm 0.028$ \\
& & & $1.0$ & $0.915 \pm 0.009$ & $0.917 \pm 0.017$ & $0.891 \pm 0.012$ & $0.903 \pm 0.019$ \\
\midrule
\multirow{9}{*}{$0.90$} & \multirow{9}{*}{$1000$} & \multirow{3}{*}{$0.0$} & $0.0$ & $0.939 \pm 0.011$ & $0.936 \pm 0.021$ & $0.935 \pm 0.010$ & $0.935 \pm 0.016$ \\
& & & $0.5$ & $0.931 \pm 0.024$ & $0.930 \pm 0.024$ & $0.925 \pm 0.019$ & $0.929 \pm 0.019$ \\
& & & $1.0$ & $0.932 \pm 0.017$ & $0.937 \pm 0.020$ & $0.925 \pm 0.014$ & $0.931 \pm 0.017$ \\
& & \multirow{3}{*}{$0.5$} & $0.0$ & $0.935 \pm 0.012$ & $0.921 \pm 0.029$ & $0.931 \pm 0.008$ & $0.929 \pm 0.018$ \\
& & & $0.5$ & $0.930 \pm 0.020$ & $0.927 \pm 0.009$ & $0.933 \pm 0.010$ & $0.923 \pm 0.010$ \\
& & & $1.0$ & $0.913 \pm 0.017$ & $0.908 \pm 0.024$ & $0.901 \pm 0.016$ & $0.901 \pm 0.013$ \\
& & \multirow{3}{*}{$1.0$} & $0.0$ & $0.928 \pm 0.011$ & $0.902 \pm 0.019$ & $0.923 \pm 0.014$ & $0.909 \pm 0.014$ \\
& & & $0.5$ & $0.916 \pm 0.024$ & $0.921 \pm 0.023$ & $0.909 \pm 0.020$ & $0.917 \pm 0.017$ \\
& & & $1.0$ & $0.927 \pm 0.015$ & $0.923 \pm 0.011$ & $0.925 \pm 0.010$ & $0.927 \pm 0.010$ \\
\midrule
\multirow{9}{*}{$0.90$} & \multirow{9}{*}{$1500$} & \multirow{3}{*}{$0.0$} & $0.0$ & $0.939 \pm 0.009$ & $0.930 \pm 0.026$ & $0.937 \pm 0.008$ & $0.931 \pm 0.017$ \\
& & & $0.5$ & $0.935 \pm 0.019$ & $0.924 \pm 0.027$ & $0.937 \pm 0.012$ & $0.921 \pm 0.017$ \\
& & & $1.0$ & $0.935 \pm 0.016$ & $0.938 \pm 0.016$ & $0.933 \pm 0.012$ & $0.933 \pm 0.012$ \\
& & \multirow{3}{*}{$0.5$} & $0.0$ & $0.928 \pm 0.022$ & $0.926 \pm 0.027$ & $0.929 \pm 0.017$ & $0.933 \pm 0.018$ \\
& & & $0.5$ & $0.934 \pm 0.010$ & $0.926 \pm 0.022$ & $0.927 \pm 0.015$ & $0.917 \pm 0.022$ \\
& & & $1.0$ & $0.925 \pm 0.018$ & $0.918 \pm 0.026$ & $0.921 \pm 0.023$ & $0.919 \pm 0.027$ \\
& & \multirow{3}{*}{$1.0$} & $0.0$ & $0.926 \pm 0.020$ & $0.927 \pm 0.022$ & $0.915 \pm 0.021$ & $0.921 \pm 0.022$ \\
& & & $0.5$ & $0.926 \pm 0.010$ & $0.921 \pm 0.011$ & $0.925 \pm 0.015$ & $0.911 \pm 0.016$ \\
& & & $1.0$ & $0.936 \pm 0.015$ & $0.931 \pm 0.013$ & $0.941 \pm 0.010$ & $0.935 \pm 0.014$ \\
\midrule
\multirow{9}{*}{$0.95$} & \multirow{9}{*}{$500$} & \multirow{3}{*}{$0.0$} & $0.0$ & $0.940 \pm 0.013$ & $0.935 \pm 0.012$ & $0.939 \pm 0.018$ & $0.933 \pm 0.014$ \\
& & & $0.5$ & $0.938 \pm 0.013$ & $0.936 \pm 0.019$ & $0.937 \pm 0.016$ & $0.935 \pm 0.015$ \\
& & & $1.0$ & $0.937 \pm 0.019$ & $0.932 \pm 0.017$ & $0.927 \pm 0.026$ & $0.925 \pm 0.024$ \\
& & \multirow{3}{*}{$0.5$} & $0.0$ & $0.930 \pm 0.011$ & $0.934 \pm 0.017$ & $0.921 \pm 0.022$ & $0.925 \pm 0.023$ \\
& & & $0.5$ & $0.937 \pm 0.013$ & $0.934 \pm 0.008$ & $0.929 \pm 0.026$ & $0.929 \pm 0.018$ \\
& & & $1.0$ & $0.936 \pm 0.020$ & $0.929 \pm 0.034$ & $0.937 \pm 0.016$ & $0.923 \pm 0.032$ \\
& & \multirow{3}{*}{$1.0$} & $0.0$ & $0.939 \pm 0.007$ & $0.938 \pm 0.012$ & $0.933 \pm 0.008$ & $0.935 \pm 0.020$ \\
& & & $0.5$ & $0.931 \pm 0.020$ & $0.929 \pm 0.018$ & $0.925 \pm 0.024$ & $0.933 \pm 0.023$ \\
& & & $1.0$ & $0.915 \pm 0.018$ & $0.921 \pm 0.011$ & $0.903 \pm 0.019$ & $0.913 \pm 0.012$ \\
\midrule
\multirow{9}{*}{$0.95$} & \multirow{9}{*}{$1000$} & \multirow{3}{*}{$0.0$} & $0.0$ & $0.937 \pm 0.013$ & $0.935 \pm 0.018$ & $0.927 \pm 0.022$ & $0.929 \pm 0.017$ \\
& & & $0.5$ & $0.937 \pm 0.013$ & $0.935 \pm 0.019$ & $0.933 \pm 0.021$ & $0.935 \pm 0.016$ \\
& & & $1.0$ & $0.941 \pm 0.008$ & $0.941 \pm 0.013$ & $0.937 \pm 0.008$ & $0.945 \pm 0.014$ \\
& & \multirow{3}{*}{$0.5$} & $0.0$ & $0.918 \pm 0.016$ & $0.919 \pm 0.018$ & $0.905 \pm 0.022$ & $0.907 \pm 0.013$ \\
& & & $0.5$ & $0.935 \pm 0.010$ & $0.930 \pm 0.014$ & $0.929 \pm 0.011$ & $0.929 \pm 0.021$ \\
& & & $1.0$ & $0.932 \pm 0.008$ & $0.930 \pm 0.016$ & $0.927 \pm 0.016$ & $0.931 \pm 0.016$ \\
& & \multirow{3}{*}{$1.0$} & $0.0$ & $0.916 \pm 0.014$ & $0.917 \pm 0.029$ & $0.903 \pm 0.015$ & $0.907 \pm 0.027$ \\
& & & $0.5$ & $0.923 \pm 0.018$ & $0.926 \pm 0.022$ & $0.925 \pm 0.015$ & $0.929 \pm 0.014$ \\
& & & $1.0$ & $0.929 \pm 0.018$ & $0.931 \pm 0.012$ & $0.913 \pm 0.021$ & $0.931 \pm 0.010$ \\
\midrule
\multirow{9}{*}{$0.95$} & \multirow{9}{*}{$1500$} & \multirow{3}{*}{$0.0$} & $0.0$ & $0.936 \pm 0.017$ & $0.941 \pm 0.016$ & $0.925 \pm 0.028$ & $0.937 \pm 0.017$ \\
& & & $0.5$ & $0.942 \pm 0.010$ & $0.932 \pm 0.024$ & $0.937 \pm 0.013$ & $0.931 \pm 0.022$ \\
& & & $1.0$ & $0.942 \pm 0.008$ & $0.932 \pm 0.015$ & $0.933 \pm 0.012$ & $0.929 \pm 0.013$ \\
& & \multirow{3}{*}{$0.5$} & $0.0$ & $0.935 \pm 0.022$ & $0.923 \pm 0.029$ & $0.931 \pm 0.020$ & $0.931 \pm 0.008$ \\
& & & $0.5$ & $0.937 \pm 0.015$ & $0.934 \pm 0.016$ & $0.933 \pm 0.016$ & $0.931 \pm 0.013$ \\
& & & $1.0$ & $0.931 \pm 0.009$ & $0.918 \pm 0.016$ & $0.927 \pm 0.017$ & $0.919 \pm 0.017$ \\
& & \multirow{3}{*}{$1.0$} & $0.0$ & $0.936 \pm 0.013$ & $0.932 \pm 0.013$ & $0.929 \pm 0.013$ & $0.929 \pm 0.014$ \\
& & & $0.5$ & $0.917 \pm 0.011$ & $0.917 \pm 0.012$ & $0.899 \pm 0.016$ & $0.909 \pm 0.014$ \\
& & & $1.0$ & $0.935 \pm 0.016$ & $0.929 \pm 0.019$ & $0.927 \pm 0.020$ & $0.929 \pm 0.020$ \\
\bottomrule
\end{tabular}
\end{table}

\clearpage

\subsection{Hyperparameter Search for GAIL}
\label{subsec:gail-hyper-search}

In this section, we report the hyperparameters used to infer user policies via Generative Adversarial Imitation Learning (GAIL), as detailed in Table~\ref{tab:gail-hyperparams}. The corresponding hyperparameter search results are summarized in Table~\ref{tab:gail-hyper-search}, where we use macro $F_1$-scores from both Random Forest (RF) and eXtreme Gradient Boosting (XGBoost) classifiers to guide optimal model selection.
\begin{table}[htbp]
\caption{Hyperparameters and parameters for GAIL experiments. $\vert\mathcal{S}\vert$ and $\vert\mathcal{A}\vert$ denote the dimensions of one-hot state and action vectors, respectively. Each stream maps to width $d_h$ before concatenation, so the combined input has size $2d_h$ regardless of $\vert\mathcal{S}\vert \neq \vert\mathcal{A}\vert$.}
\small
\begin{tabular}{lc}
\toprule
\textbf{Hyperparameter} & \textbf{Values} \\
\midrule
Learning rate & $3 \times 10^{-4}$ \\
Batch size & 64 \\
Steps per rollout ($n_{\text{steps}}$) & 128 \\
Entropy coefficient ($c_{\text{ent}}$) & 0.01 \\
Discount factor ($\gamma$) & $\{0.95, 0.99\}$ \\
Policy network structure & (16, 16) \\
Value network structure & (16, 16) \\
Discriminator updates / round & 5 \\
Reward net structure & Obs: ($d_s$, $d_h$), Act: ($d_a$, $d_h$), Combined: ($2d_h$, $d_h$, 1) \\
Reward net hidden nodes ($d_h$) & $\{4, 8, 16\}$ \\
Total training steps & $\{5{,}000,\;10{,}000,\;20{,}000\}$ \\
Running norm (RN) & $\{0,1\}$ \\
Optimizer & Adam\\
\bottomrule
\end{tabular}
\label{tab:gail-hyperparams}
\end{table}

\begin{table}[htbp]
\caption{GAIL Classification Performance.}
\label{tab:gail-hyper-search}
\scriptsize
\begin{tabular}{rrrrcccc}
\toprule
 & & & & \multicolumn{2}{c}{$F_1$-Score} & \multicolumn{2}{c}{Recall} \\ \cmidrule(lr){5-6} \cmidrule(lr){7-8}
$\gamma$ & $d_h$ & Steps & RN & RF & XGBoost & RF & XGBoost \\
\midrule
\multirow{6}{*}{$0.95$} & \multirow{6}{*}{$4$} & \multirow{2}{*}{$5,000$} & $0$ & $0.948 \pm 0.015$ & $0.939 \pm 0.019$ & $0.935 \pm 0.012$ & $0.927 \pm 0.017$ \\
& & & $1$ & $0.922 \pm 0.025$ & $0.920 \pm 0.014$ & $0.925 \pm 0.029$ & $0.925 \pm 0.023$ \\
& & \multirow{2}{*}{$10,000$} & $0$ & $0.928 \pm 0.018$ & $0.924 \pm 0.018$ & $0.933 \pm 0.023$ & $0.933 \pm 0.012$ \\
& & & $1$ & $0.936 \pm 0.018$ & $0.932 \pm 0.012$ & $0.931 \pm 0.022$ & $0.927 \pm 0.008$ \\
& & \multirow{2}{*}{$20,000$} & $0$ & $0.901 \pm 0.015$ & $0.893 \pm 0.008$ & $0.889 \pm 0.030$ & $0.885 \pm 0.015$ \\
& & & $1$ & $0.933 \pm 0.019$ & $0.923 \pm 0.024$ & $0.931 \pm 0.017$ & $0.919 \pm 0.019$ \\
\midrule
\multirow{6}{*}{$0.95$} & \multirow{6}{*}{$8$} & \multirow{2}{*}{$5,000$} & $0$ & $0.928 \pm 0.016$ & $0.936 \pm 0.024$ & $0.917 \pm 0.026$ & $0.933 \pm 0.031$ \\
& & & $1$ & $0.946 \pm 0.017$ & $0.939 \pm 0.011$ & $0.939 \pm 0.016$ & $0.943 \pm 0.017$ \\
& & \multirow{2}{*}{$10,000$} & $0$ & $0.924 \pm 0.023$ & $0.917 \pm 0.017$ & $0.923 \pm 0.025$ & $0.909 \pm 0.019$ \\
& & & $1$ & $0.926 \pm 0.028$ & $0.940 \pm 0.014$ & $0.921 \pm 0.031$ & $0.943 \pm 0.014$ \\
& & \multirow{2}{*}{$20,000$} & $0$ & $0.924 \pm 0.012$ & $0.917 \pm 0.018$ & $0.935 \pm 0.012$ & $0.917 \pm 0.025$ \\
& & & $1$ & $0.936 \pm 0.019$ & $0.922 \pm 0.019$ & $0.937 \pm 0.023$ & $0.937 \pm 0.019$ \\
\midrule
\multirow{6}{*}{$0.95$} & \multirow{6}{*}{$16$} & \multirow{2}{*}{$5,000$} & $0$ & $0.934 \pm 0.015$ & $0.932 \pm 0.019$ & $0.925 \pm 0.015$ & $0.923 \pm 0.028$ \\
& & & $1$ & $0.945 \pm 0.013$ & $0.936 \pm 0.010$ & $0.939 \pm 0.020$ & $0.935 \pm 0.018$ \\
& & \multirow{2}{*}{$10,000$} & $0$ & $0.933 \pm 0.015$ & $0.931 \pm 0.017$ & $0.933 \pm 0.012$ & $0.935 \pm 0.023$ \\
& & & $1$ & $0.940 \pm 0.013$ & $0.931 \pm 0.019$ & $0.952 \pm 0.015$ & $0.933 \pm 0.017$ \\
& & \multirow{2}{*}{$20,000$} & $0$ & $0.943 \pm 0.013$ & $0.930 \pm 0.014$ & $0.956 \pm 0.015$ & $0.941 \pm 0.019$ \\
& & & $1$ & $0.936 \pm 0.017$ & $0.932 \pm 0.018$ & $0.939 \pm 0.016$ & $0.937 \pm 0.019$ \\
\midrule
\multirow{6}{*}{$0.99$} & \multirow{6}{*}{$4$} & \multirow{2}{*}{$5,000$} & $0$ & $0.927 \pm 0.013$ & $0.926 \pm 0.013$ & $0.909 \pm 0.010$ & $0.915 \pm 0.009$ \\
& & & $1$ & $0.928 \pm 0.022$ & $0.929 \pm 0.013$ & $0.907 \pm 0.019$ & $0.909 \pm 0.020$ \\
& & \multirow{2}{*}{$10,000$} & $0$ & $0.945 \pm 0.021$ & $0.946 \pm 0.012$ & $0.943 \pm 0.018$ & $0.956 \pm 0.014$ \\
 & & & $1$ & \textbf{\boldmath $0.956 \pm 0.015$} & $0.939 \pm 0.008$ & $0.956 \pm 0.015$ & $0.947 \pm 0.019$ \\
& & \multirow{2}{*}{$20,000$} & $0$ & $0.923 \pm 0.025$ & $0.928 \pm 0.022$ & $0.917 \pm 0.035$ & $0.925 \pm 0.030$ \\
& & & $1$ & $0.937 \pm 0.025$ & $0.931 \pm 0.012$ & $0.947 \pm 0.027$ & $0.943 \pm 0.020$ \\
\midrule
\multirow{6}{*}{$0.99$} & \multirow{6}{*}{$8$} & \multirow{2}{*}{$5,000$} & $0$ & $0.933 \pm 0.014$ & $0.930 \pm 0.019$ & $0.911 \pm 0.015$ & $0.919 \pm 0.020$ \\
& & & $1$ & $0.954 \pm 0.020$ & $0.956 \pm 0.017$ & $0.954 \pm 0.021$ & $0.960 \pm 0.024$ \\
& & \multirow{2}{*}{$10,000$} & $0$ & $0.940 \pm 0.019$ & $0.933 \pm 0.021$ & $0.929 \pm 0.020$ & $0.937 \pm 0.024$ \\
& & & $1$ & $0.946 \pm 0.011$ & $0.946 \pm 0.020$ & $0.943 \pm 0.018$ & $0.954 \pm 0.015$ \\
& & \multirow{2}{*}{$20,000$} & $0$ & $0.938 \pm 0.013$ & $0.934 \pm 0.019$ & $0.935 \pm 0.012$ & $0.933 \pm 0.021$ \\
& & & $1$ & $0.938 \pm 0.010$ & $0.937 \pm 0.014$ & $0.943 \pm 0.009$ & $0.941 \pm 0.011$ \\
\midrule
\multirow{6}{*}{$0.99$} & \multirow{6}{*}{$16$} & \multirow{2}{*}{$5,000$} & $0$ & $0.935 \pm 0.013$ & $0.928 \pm 0.014$ & $0.923 \pm 0.017$ & $0.925 \pm 0.021$ \\
& & & $1$ & $0.940 \pm 0.017$ & $0.942 \pm 0.018$ & $0.943 \pm 0.014$ & $0.954 \pm 0.023$ \\
& & \multirow{2}{*}{$10,000$} & $0$ & $0.948 \pm 0.012$ & $0.949 \pm 0.004$ & $0.949 \pm 0.014$ & $0.954 \pm 0.015$ \\
& & & $1$ & $0.943 \pm 0.016$ & $0.938 \pm 0.019$ & $0.941 \pm 0.021$ & $0.943 \pm 0.023$ \\
& & \multirow{2}{*}{$20,000$} & $0$ & $0.942 \pm 0.012$ & $0.932 \pm 0.015$ & $0.941 \pm 0.018$ & $0.927 \pm 0.017$ \\
& & & $1$ & $0.937 \pm 0.014$ & $0.940 \pm 0.019$ & $0.941 \pm 0.022$ & $0.952 \pm 0.017$ \\
\bottomrule
\end{tabular}
\end{table}

\clearpage
\bibliographystyle{abbrev_custom}
\bibliography{references}

\end{document}